\documentclass[a4paper,12pt]{article}

\usepackage{type1cm}        
\usepackage{makeidx}         
\usepackage{graphicx}        
\usepackage{multicol}        
\usepackage[bottom]{footmisc}

\usepackage{newtxtext}       %
\usepackage[varvw]{newtxmath}       
\usepackage{amsmath}

\makeindex             

\usepackage{subcaption}
\usepackage{hyperref}
\usepackage{cite}
\newcommand{\etal}{\textit{et al.}}

\newcommand{\aprio}{\textit{a-priori }}
\newcommand{\apost}{\textit{a-posteriori }}

\usepackage[margin=1in]{geometry}

\usepackage{authblk}
\usepackage{xcolor}
\usepackage{fancyhdr}
\fancypagestyle{plain}{%
	\fancyhf{}%
	\fancyhead[C]{\textcolor{red}{Accepted Manuscript: This version has been accepted and will appear in High Performance Computing in Science and Engineering '24.}}%
}

\title{Super-resolution of turbulent velocity and scalar fields using different scalar distributions}
\author[1]{Ali Shamooni\thanks{\texttt{ali.shamooni@irst.uni-stuttgart.de}}}
\author[2]{Oliver T. Stein\thanks{\texttt{oliver.t.stein@kit.edu}}}
\author[3]{Andreas Kronenburg\thanks{\texttt{andreas.kronenburg@irst.uni-stuttgart.de}}}
\affil[1]{Institute for Reactive Flows (IRST), University of Stuttgart,  Pfaffenwaldring 31, 70569 Stuttgart, Germany}
\affil[2]{Engler-Bunte-Institut, Simulation of Reacting Thermo-Fluid Systems, Karlsruhe Institute for Technology, Engler-Bunte-Ring 7, 76131 Karlsruhe, Germany}
\affil[3]{Institute for Reactive Flows (IRST), University of Stuttgart,  Pfaffenwaldring 31, 70569 Stuttgart, Germany}
\date{}

\begin{document}
	
	\maketitle
	
	\abstract{
		In recent years, sub-grid models for turbulent mixing  have been developed by data-driven methods for large eddy simulation (LES). Super-resolution is a data-driven deconvolution technique in which deep convolutional neural networks are trained using direct numerical simulation (DNS) data to learn  mappings between the input data from a low resolution  domain to the super-resolved high resolution output domain. 
		While the technique has been of  a great success in \textit{a-priori} tests, the assessment of its generalization capabilities is required for further \textit{a-posteriori} applications.  
		In this study we assess the generalization capability of a super-resolution generative adversarial network (GAN) in reconstructing scalars with different distributions. 
		Forced turbulence mixing DNS data with a fixed Reynolds number but different bulk scalar distributions, are generated and used as training/testing datasets.  
		The results show that  the velocity vector field can be reconstructed well, but the model fails to super-resolve the scalars from out-of-sample  distributions. 
		Including  two extreme mixture fraction distributions, namely double Pareto and semi-Gaussian,  in the training dataset significantly improves the performance of the model, not only for those distributions, but also for previously unseen bimodal distributions.   }

	\section{Introduction}
	\label{sec:2}
	
	Turbulent scalar mixing occurs in many engineering applications such as most combustion systems of engineering interest. 
	While processes such as convection and diffusion with different scales are involved in turbulent mixing, the  mixing occurs mainly at the smallest flow scales~\cite{SreenivasanK2019_PNASUSA} and thus, an accurate representation of these scales is of great importance from a modeling point of view. 
	DNS in which all turbulent scales  are resolved is still computationally prohibitive and LES is more common. The governing equations in LES have unclosed terms that originate from filtering the non-linear terms in the Navier-Stokes equations, and they require closure. 
	Deconvolution models have been developed to approximate an inverse to the LES filtering operator and to reconstruct the LES sub-grid information explicitly from the deconvolved variables~\cite{DomingoP2017_CF,WangQ2019_CF}. 
	
	Recently, deep convolutional neural networks (CNNs) and transformer-based networks  have been used for modeling of sub-grid scale stresses and scalar  fluxes~\cite{LapeyreC2019_CF,KimJ2020_JFM,RenJ2021_PF,BodeM2021_PCI,GaudingM2021_CONF,GrengaT2022_CST,XuQ2023_POF,NistaL2023_PCI,PangZ2024_CF}. Very deep CNNs such as residual-in-residual blocks (RRDBs) with adversarial training using generative adversarial networks (GANs)~\cite{GoodfellowI2020_CA} have shown superior  results~\cite{KimJ2020_JFM,BodeM2021_PCI,GaudingM2021_CONF,GrengaT2022_CST,NistaL2023_PCI}. 
	Generally, in CNNs each layer is a set of non-linear functions of weighted sums from 
	spatially close subsets of outputs from the previous layer, which allows for capturing spatial structures (correlations) in the training data. 
	
	Specifically in scalar mixing modeling, Bode \etal~\cite{BodeM2021_PCI} used GAN with an augmented physical loss function viz. gradient and divergence of velocity components losses (PIERSGAN) to perform super-resolution  (SR) of  velocity and scalar fields from filtered DNS data. 
	The authors trained and tested the network using a database of decaying homogeneous isotropic turbulence (HIT) with a forced scalar field using a gradient scalar forcing method. 
	Gauding and Bode~\cite{GaudingM2021_CONF} expanded the applicability of PIERSGAN by training it on a decaying  HIT database as well as on coarse-grained jet flow data. This enabled  them to test the model for SR of a mixture fraction field in a turbulent jet flow. 
	It should be noted that while the scalar probability density function (PDF) is different from their previous work~\cite{BodeM2021_PCI}, the training and testing datasets still have similar scalar PDFs.  Nista \etal~\cite{NistaL2023_PCI} studied the generalization capability of a similar network (without divergence loss)  on different Reynolds and Karlovitz numbers between training/testing datasets when applying the network on SR of velocity and scalar fields of jet flame data.  
	They demonstrated that the ratio of filter width to Kolmogorov scale must be preserved between training and testing data to ensure the generalization of the network to accurately reconstruct the sub-filter stress tensor and scalar fluxes. They also called for further investigations since the performance of SR models to out-of-sample conditions is still widely unexplored. 
	This is a necessary step toward \apost application of SR as sub-grid turbulence models. 
	To the authors' knowledge, there is no comprehensive study that investigates the effect of scalar distributions on the generalization capability of the trained SR networks. 
	Of particular interest is the distribution of mixture fraction which is a passive scalar and characterizes mixing between fuel and oxidizer in the combustion chamber. 
	In  \apost applications like e.g. jet flames, different mixture fraction PDFs may occur aas their exact shape will depend on both, the specific configuration (e.g. free jet, bluff body or swirl) and the exact location within the setup. 
	Stratified flames may serve as an example as they  show an extreme double Pareto mixture fraction distribution   near the  entrance regions, while further downstream Gaussian distributions are observed due to perfect mixing. 
	As such, it is important to  investigate the performance of super-resolution networks in such scenarios.  
	
	In the current study we aim to address the above-mentioned gap by exploring the  generalization capability of a super-resolution model to different scalar distributions. To this end, we generate DNS databases of forced turbulence  with different  mixture faction  distributions. The DNS data  is employed as training/testing datasets for super-resolution of filtered DNS mixture fraction fields in an \aprio manner.

	\section{Methodology}
	The closure problem in LES can be formulated as  $\overline{\mathcal{N}}(\mathcal{Q})\neq\mathcal{N}(\overline{\mathcal{Q}})$, where $\mathcal{Q}$ denotes vector/scalar primitive variables, $\overline{(.)}$ implies Reynolds filtering, and  $\mathcal{N}$ is a corresponding  non-linear operator such as a chemical source term or a dissipation rate. $\mathcal{Q}$ is unresolved and only the filtered value $\overline{\mathcal{Q}}$ is known.
	For incompressible turbulent mixing considered in the current study, the variables are velocity vector components $u_i$ and the mixture fraction field $Z$ where the non-linear operators contain the turbulent stress fields  and the turbulent scalar fluxes.  
	The deconvolution task can then be formulated as: given $\overline{\mathcal{Q}}$, approximate ${\mathcal{Q}}^*$ such that ${\mathcal{Q}}^*\simeq{\mathcal{Q}}$. 
	In such a case, $\overline{\mathcal{N}}(\mathcal{Q})$ can be approximated by  $\overline{\mathcal{N}}(\mathcal{Q}^*)$. In computer vision this is called super-resolution  (SR).  
	
	Deep CNNs such as residual-in-residual dense blocks (RRDBs) with GAN training~\cite{WangX2019_}  are state-of-the-art DL  approaches  to deconvolve velocity vectors and mixture fraction~\cite{KimH2020_JFM,BodeM2021_PCI,ChungW2023_CONF}. 
	In GANs, two competing CNN networks are used, where the \textit{generator} network learns to generate fake outputs resembling real data and the \textit{discriminator} network learns to distinguish between fake and real data which in turn forces the generator to improve its output to keep fooling the discriminator. Eventually, this process reaches an equilibrium in which the generated data is indistinguishable from real data. 
	Pure CNNs may give better pixel-wise (lower mean-square error) results than GANs, however, high wave-number structures are represented better with GANs~\cite{WangX2019_,KimH2020_JFM}. 
	In such techniques deep networks are trained by using high resolution data ($\textbf{q}_{HR}\equiv {\mathcal{Q}}$) to map  the low resolution data ($\textbf{q}_{LR}\equiv \overline{\mathcal{Q}}$) to super-resolved fields ($\textbf{q}_{SR}\equiv {\mathcal{Q}^*}$). 
	Here, we also adopt the RRDB network for the SR task.

	The network is composed of 23 RRDB blocks, each block with 3 residual dense blocks (RDBs). 
	The RDBs are composed of five 2D convolutional layers  with kernel size of 3 and stride of 1. 
	The input channel size of each  convolutional layer of RDB  is 64 and increased by a growth rate of 32 after each layer. 
	The output channels size is 32 for all layers in RDB with the exception of the last one which is 64. 
	After each RDB, a leaky ReLu activation function is imposed. 
	Skip connections  are employed between RRDBs and within each RDB with a mixing factor of 0.2. 
	Two up-sampling layers with fixed channel size of 64 are placed at the end of the network before the last convolutional layer with 4 output channels i.e. velocity components and the mixture fraction. 
	The up-sampling layers are composed of a convolutional layer followed by nearest neighbor interpolation and the activation function.  
	The total number of trainable parameters of the RRDB as the SR generator is approximately 16.7 million.

	An auxiliary  discriminator network is used for GAN training which is adopted from~\cite{WangX2021_CONF}. It is a U-Net discriminator with spectral normalization regularization~\cite{MiyatoT2018_CONF,WangX2021_CONF} which has an improved stability over conventional CNN-based discriminators~\cite{WangX2019_,BodeM2021_PCI}. 
	The total number of trainable parameters of the  U-Net discriminator   is approximately 4.4 million. 
	
	The total loss function of the network is composed of pixel, physical, and adversarial  losses, 
	\begin{equation}\label{eq:Lgan}
		\mathcal{L}_{GAN} = \beta_{\mathcal{L}_{1}} \mathcal{L}_{1} + \beta_{\mathcal{L}_{grad}} \mathcal{L}_{grad} + \beta_{\mathcal{L}_{AD}} \mathcal{L}_{AD},
	\end{equation}
	where  $ \mathcal{L}_{1}=\mathbb{E}_{\textbf{q}_{LR}}|| G(\textbf{q}_{LR}) - \textbf{q}_{HR}||_1$ is the pixel-wise 1-norm distance between SR data, viz. $ \textbf{q}_{SR}=G(\textbf{q}_{LR})$, and the ground-truth high resolution (HR) data from DNS $\textbf{q}_{LR}$. Here $G$
	stands for the \textit{generator} network which is the RRDB explained above.  
	$\mathbb{E}$ is the mean over all data in each mini batch od input data. 
	The gradient loss, $\mathcal{L}_{grad}$, is a physical loss which minimizes the 2-norm distance  between the gradients of fields, viz. $ \mathcal{L}_{grad}=\mathbb{E}_{  \textbf{q}_{LR}}|| G(\nabla \textbf{q}_{LR}) - \nabla \textbf{q}_{HR}||_2  $. 
	A relativistic  adversarial loss~\cite{WangX2019_}, $\mathcal{L}_{AD}$, is considered  to update the discriminator and generator simultaneously. Note that the discriminator network only minimizes $\mathcal{L}_{AD}$. 
	Using the ground-truth and reconstructed data from the generator, the discriminator is trained in an adversarial manner to assess how close the generator outputs are  to the ground-truth. 
	The coefficients $\beta_i$ are hyper-parameters that are tuned to make the loss parts to be of the same order.

	\section{Data generation and training strategy}

	\subsection{DNS of forced turbulent mixing}
	
	The incompressible Navier-Stokes  equations and an additional passive scalar transport equation are solved in a periodic box  using the OpenFOAM software package with second-order convection and time discretization schemes and a third order interpolation scheme for the diffusion terms. Courant–Friedrichs–Lewy (CFL) number is kept below 0.1.  The data are written in HDF5 format for efficient parallel I/O performance. 
	Simulations are performed on  1024 AMD-7742 cores using   \textit{Hawk}.  
	Typical simulations require 45,000 time steps to reach stationary conditions and further 100,000 time steps for data collection, which cost approximately 36k CPUh on 8 nodes of \textit{Hawk}.
	
\begin{figure}[!htb]
		\centering
		\includegraphics[width=0.7\textwidth]{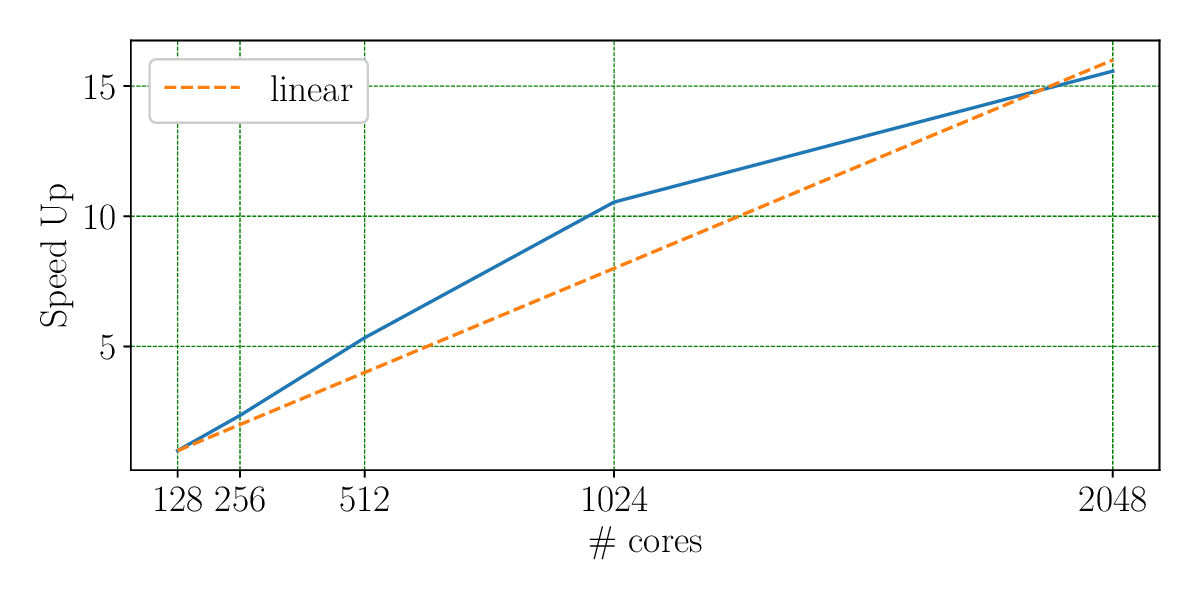}
		\caption{Strong scaling performance of the DNS code.   }
		\label{fig:strongS}
	\end{figure}

	The strong scaling performance of the solver can be seen in Fig.\,\ref{fig:strongS}. The test is carried out for a $256^3$ simulation  using 1-16 nodes of \textit{Hawk} each with 128 cores.  
	As can be seen, the code scales very well and super-linear scaling is achieved when using 1024 cores. 

	\subsection{Dataset description}
	Different forced  turbulence-in-a-box cases are generated as training/testing databases using a linear velocity forcing~\cite{BassenneM2016_PF}. The turbulent mixing of different scalar distributions has been achieved by forcing the scalar independently using a scalar forcing method which is based on~\cite{BrearleyP2022_FTC}. 
	
	Given the RMS of the velocity ($u_{rms,0}$) and a prescribed length scale ($L_{t,0}$), the velocity fields are initiated by a prescribed von Karman spectrum from different random states~\cite{SaadT2017_AJ}.
	The specifications of DNS cases have been reported in Table~\ref{tab:sharedSpecInit}. 
	\begin{table}[!h]
		
		\caption{HIT DNS setups.}
		\label{tab:sharedSpecInit}       

		\begin{center}
			\begin{tabular}{p{7.5cm}p{2cm}p{2cm}p{2cm}}
			\hline 
			Property & \multicolumn{3}{c}{value}  \\
			\hline 
			Gas kinematic viscosity $(\nu_0) [\mathrm{m^2/s}]$  & \multicolumn{3}{c}{$3.96\cdot10^{-5}$} \\
			Domain length  $\left(L_x = L_y = L_z \right) [m]$ &\multicolumn{3}{c}{0.0512} \\
			Grid    $\left(N_x = N_y = N_z \right)$ &\multicolumn{3}{c}{256} \\
			$\Delta_x/\eta$ & \multicolumn{3}{c}{ $\approx 1.57$} \\
			Taylor Reynolds number ${\mathit{Re}_{\lambda,0}}$  &\multicolumn{3}{c}{ 82}\\
			Kolmogorov time scale $\tau_{\eta,0}$ [s]  &\multicolumn{3}{c}{ 0.00025}\\
			Turbulent kinetic energy $k_0 [m^2/s^2]$  &\multicolumn{3}{c}{ 4.99} \\
			R.M.S velocity $u_{rms,0} [m/s]$ &\multicolumn{3}{c}{ 1.82} \\
			Turbulent length scale  $(L_{t,0}= {u_{rms,0}^3}/{\varepsilon_0})\, [\mathrm{m}]$  &\multicolumn{3}{c}{  0.0097} \\
			Turbulent Reynolds number ${\mathit{Re}}_{t,0} = {u_{rms,0}L_{t,0}}/{\nu} $  &  \multicolumn{3}{c}{447}\\ 
			Turbulent time scale   	${\tau_{L,0}} [s]$ &\multicolumn{3}{c}{ 0.0053}  \\
			\hline 
			& Case1   & Case2    & Case3 \\
			& (double Pareto) &   (semi-Gaussian)    &      (bimodal) \\
			R.M.S scalar $Z_{rms,0}$  & 0.415  &  0.1 &0.31    \\
			\hline 
		\end{tabular}\\
		\end{center}
	
	\end{table}
	\begin{figure}[!htb]
		\centering
		\begin{minipage}[l]{.3\textwidth}
			\centering
			\includegraphics[width=\textwidth]{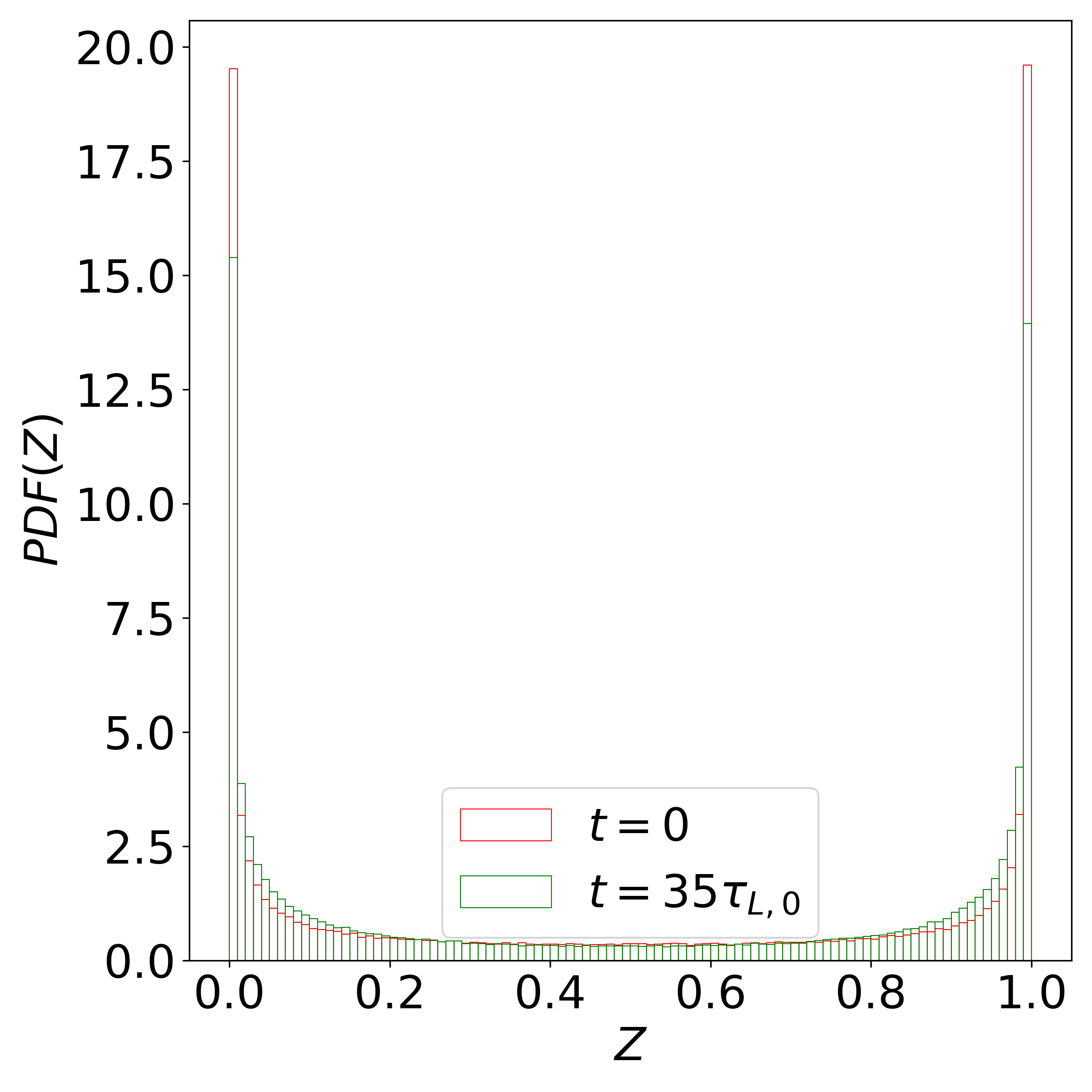}
			\subcaption{}\label{fig:f_0.19984}
		\end{minipage}
		\begin{minipage}[c]{.3\textwidth}
			\centering
			\includegraphics[width=\textwidth]{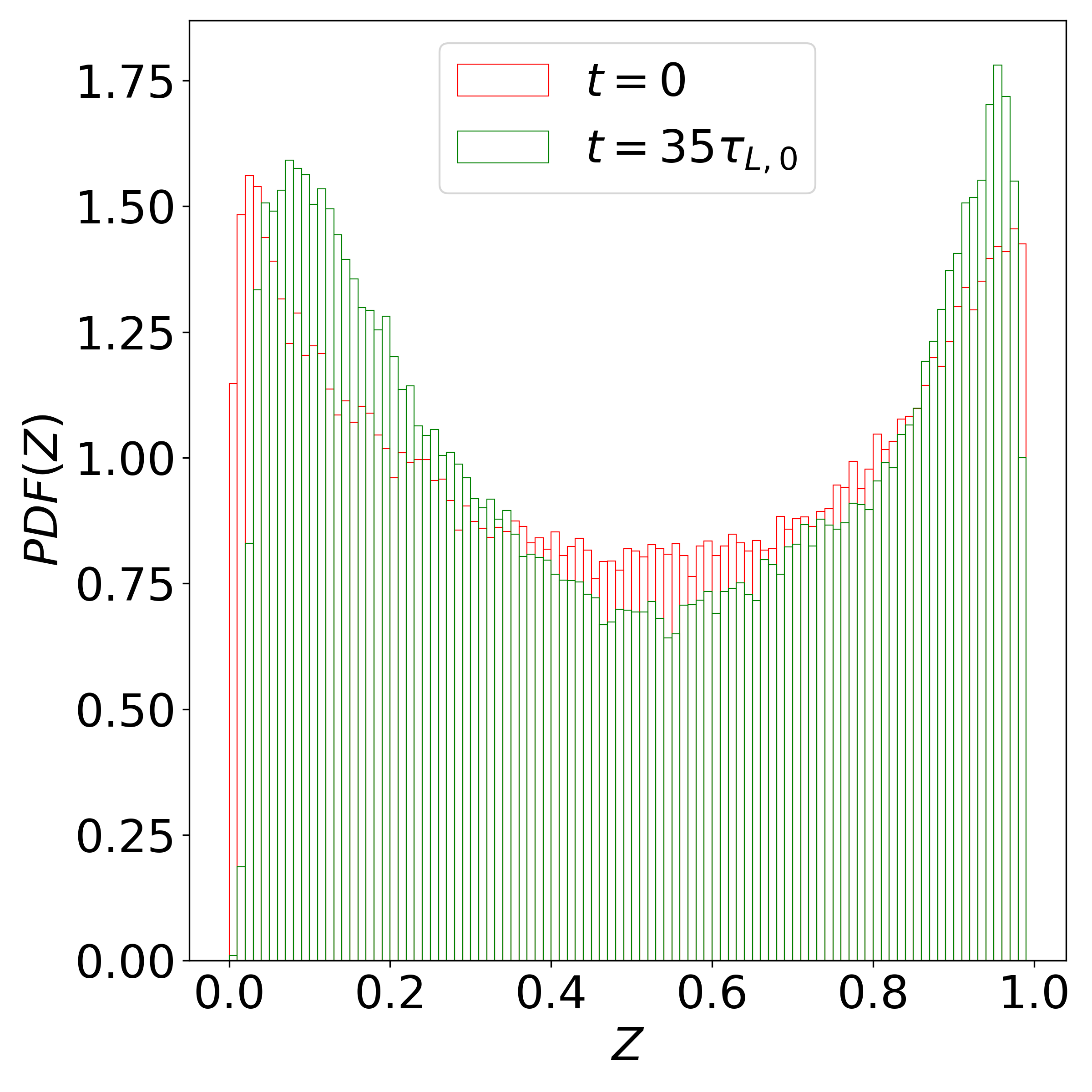}
			\subcaption{}\label{fig:f_rms031_0.19984}
		\end{minipage}  
		\begin{minipage}[r]{.3\textwidth}
			\centering
			\includegraphics[width=\textwidth]{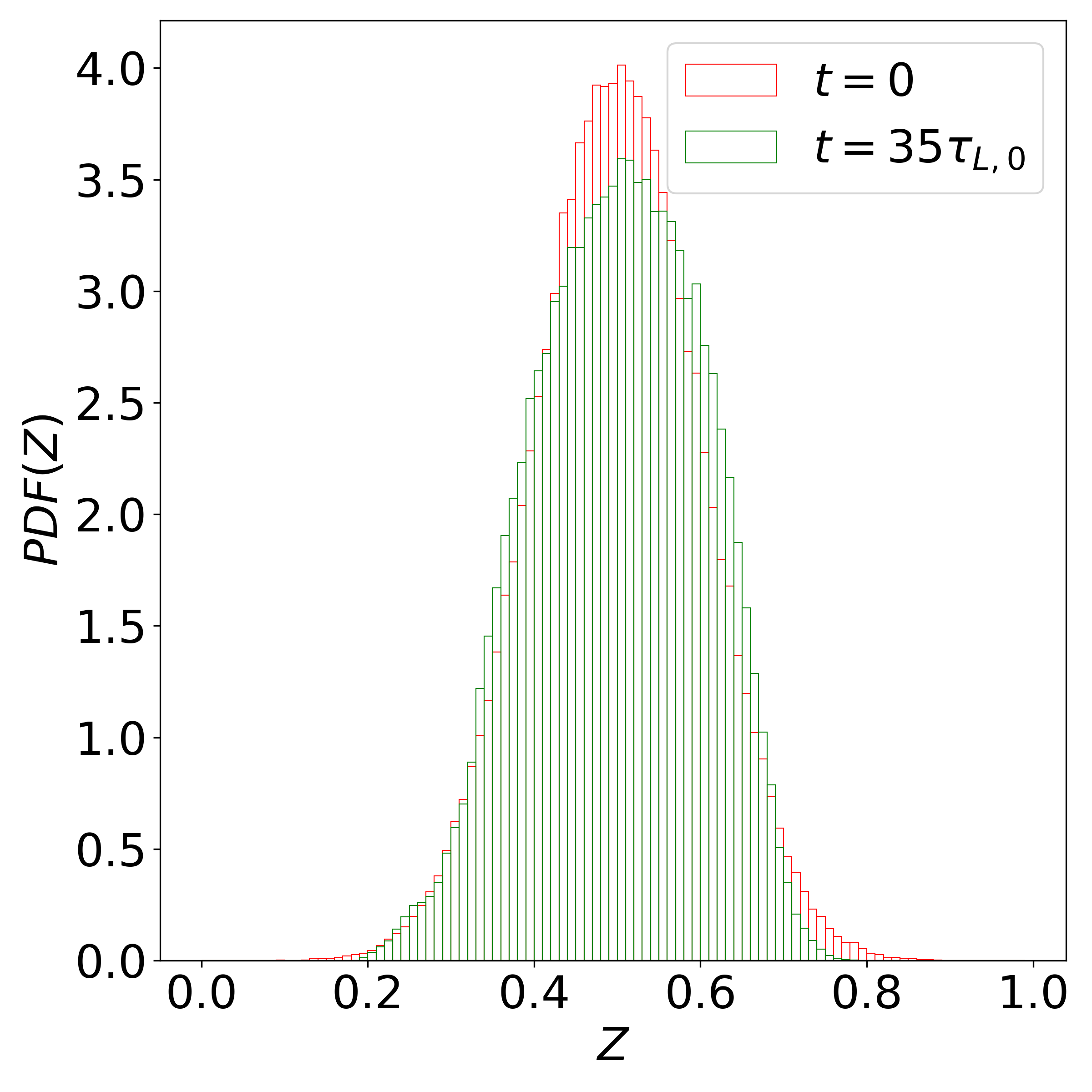}
			\subcaption{}\label{fig:f_rms010Gaussian_0.19984}
		\end{minipage}  \\
		\begin{minipage}[l]{.3\textwidth}
			\centering
			\includegraphics[height=\textwidth]{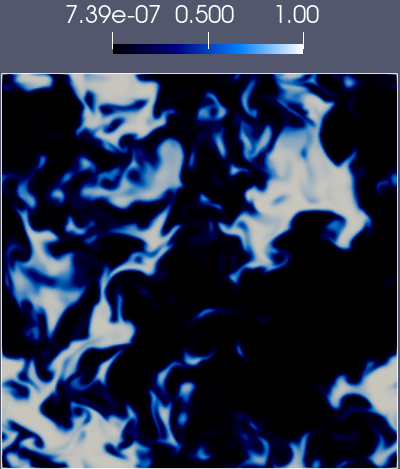}
			\subcaption{}\label{fig:f_doublePareto_2}
		\end{minipage}
		\begin{minipage}[c]{.3\textwidth}
			\centering
			\includegraphics[height=\textwidth]{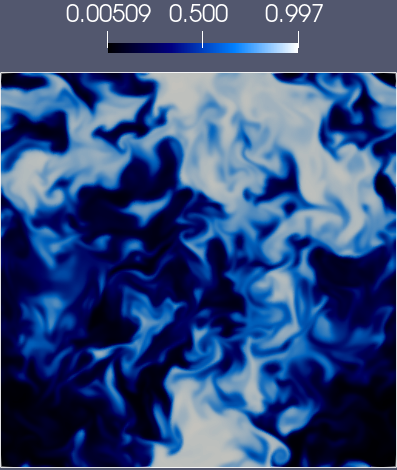}
			\subcaption{}\label{fig:f_rms_031_2}
		\end{minipage}
		\begin{minipage}[r]{.3\textwidth}
			\centering
			\includegraphics[height=\textwidth]{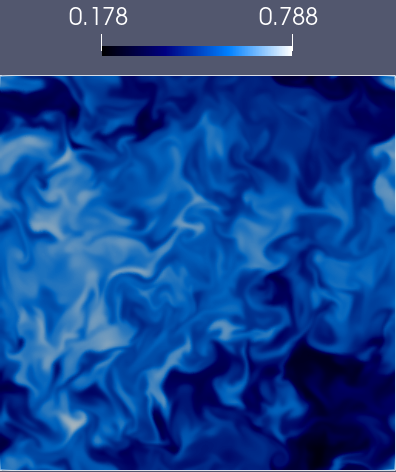}
			\subcaption{}\label{fig:f_rms_010Gaussian_2}
		\end{minipage}  
		\caption{Initial and final distribution of mixture fraction  ($Z$) for \textbf{(a)} forced double Pareto, \textbf{(b)} forced bimodal, and \textbf{(c)} forced semi-Gaussian scalars from DNS of turbulent mixing. \textbf{(d, e, f)} show final scalar fields  (at $t=36 \tau_{L,0}$)  for   \textbf{(d)}  double Pareto, \textbf{(e)}  bimodal, and \textbf{(e)}  semi-Gaussian scalar distributions. }
		\label{fig:InitialFinalZDist}
	\end{figure}
	
	The scalar fields are initiated by diffused random noise. The procedure is as follows:  Given a desired scalar length scale ($=0.5 L_{t,0}$) and mean value ($=0.5$), diffusion is applied to create ``blobs'' of mixture fraction in a pre-processing step. Different scalar distributions are then generated by applying a tangent hyperbolic transformation and letting the generated fields diffuse for different numbers of time steps. Using this procedure, three initially different scalar distributions are generated which are labeled \textit{Case1-Case3} in Table~\ref{tab:sharedSpecInit}. 
	
	In Fig.\,\ref{fig:InitialFinalZDist} the generated fields with two extreme distributions, i.e., double Pareto  and semi-Gaussian distributions as well as an intermediate bimodal distribution   at  the initial   and final ($t=36\tau_{L,0}$)  simulation times are shown. It can be seen that the employed  scalar forcing method can retain the scalar distribution shape during a long period of simulation time. It should be noted that without any forcing method, the initial scalar field diffuses  quickly  and a uniform distribution is achieved which does not provide enough training data for the training of very deep learning networks.

\begin{figure}[!htb]
		\centering
		\includegraphics[width=0.5\textwidth]{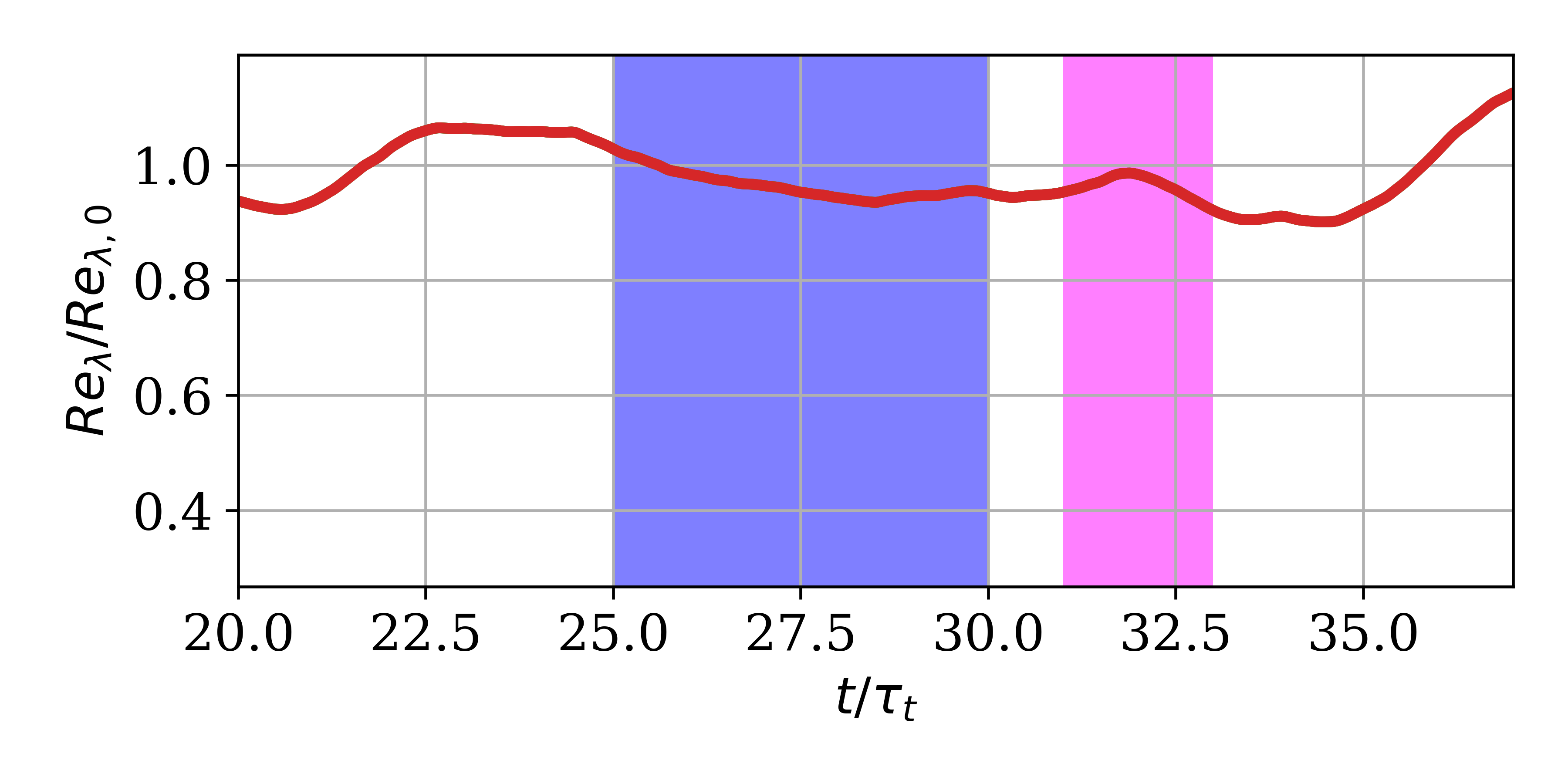}
		\caption{The evolution of Taylor Reynolds number versus time for \textit{Case1}. The blue and magenta colored backgrounds show the time period for training and testing data samplings, respectively.   }
		\label{fig:ReLambda_vs_time}
	\end{figure}

	The stationary states of the DNS cases are  detected when target quantities such as target turbulent kinetic energy or Taylor Reynolds number reach a plateau as can be seen for $Re_{\lambda}$ of \textit{Case1} in Fig.\,\ref{fig:ReLambda_vs_time}.   
	\begin{figure}[!htb]
		\centering
		\begin{minipage}[l]{.49\textwidth}
			\centering
			\includegraphics[width=\textwidth]{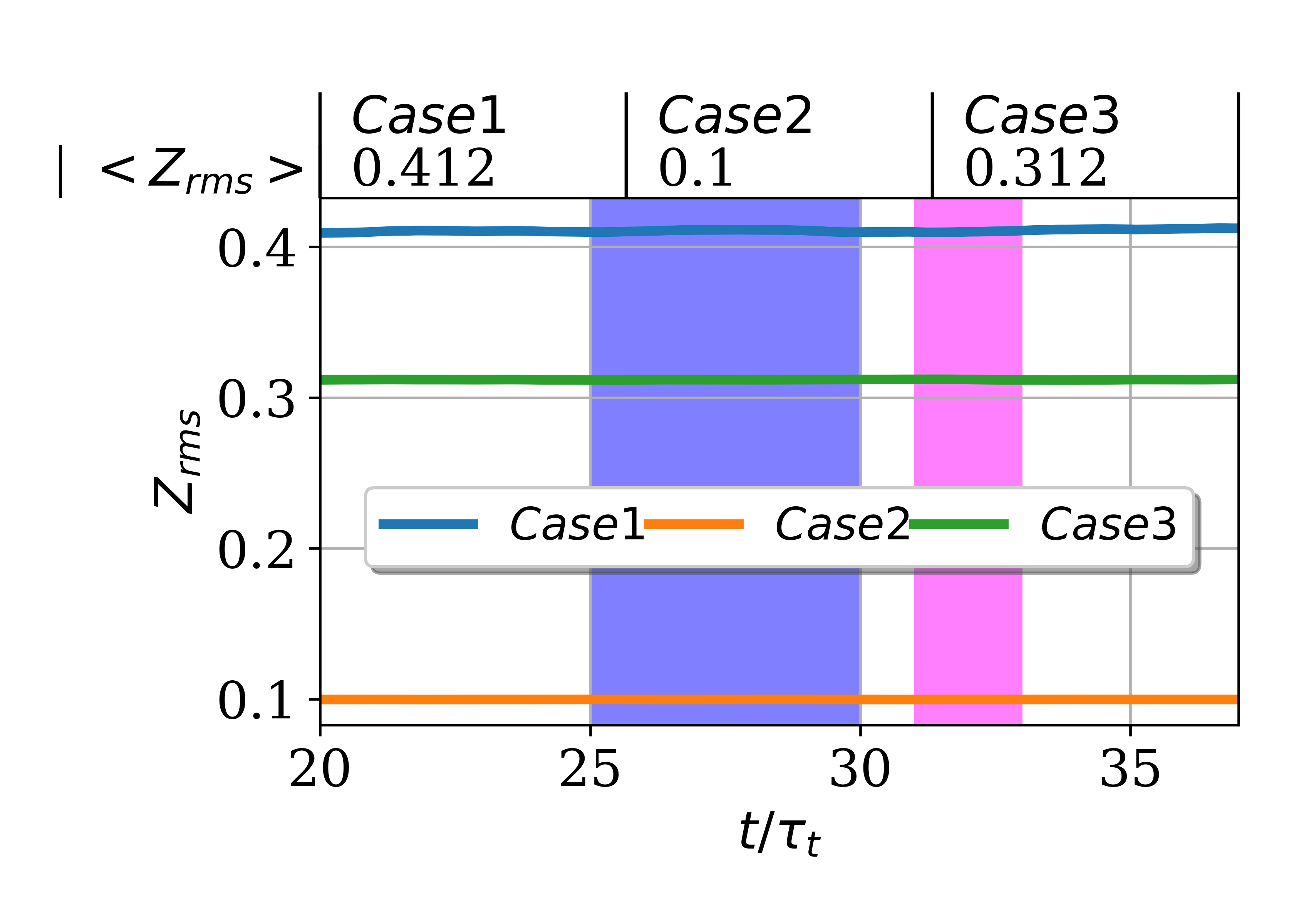}
			\label{fig:Zrms_cases}
		\end{minipage}
		\begin{minipage}[c]{.49\textwidth}
			\centering
			\includegraphics[width=\textwidth]{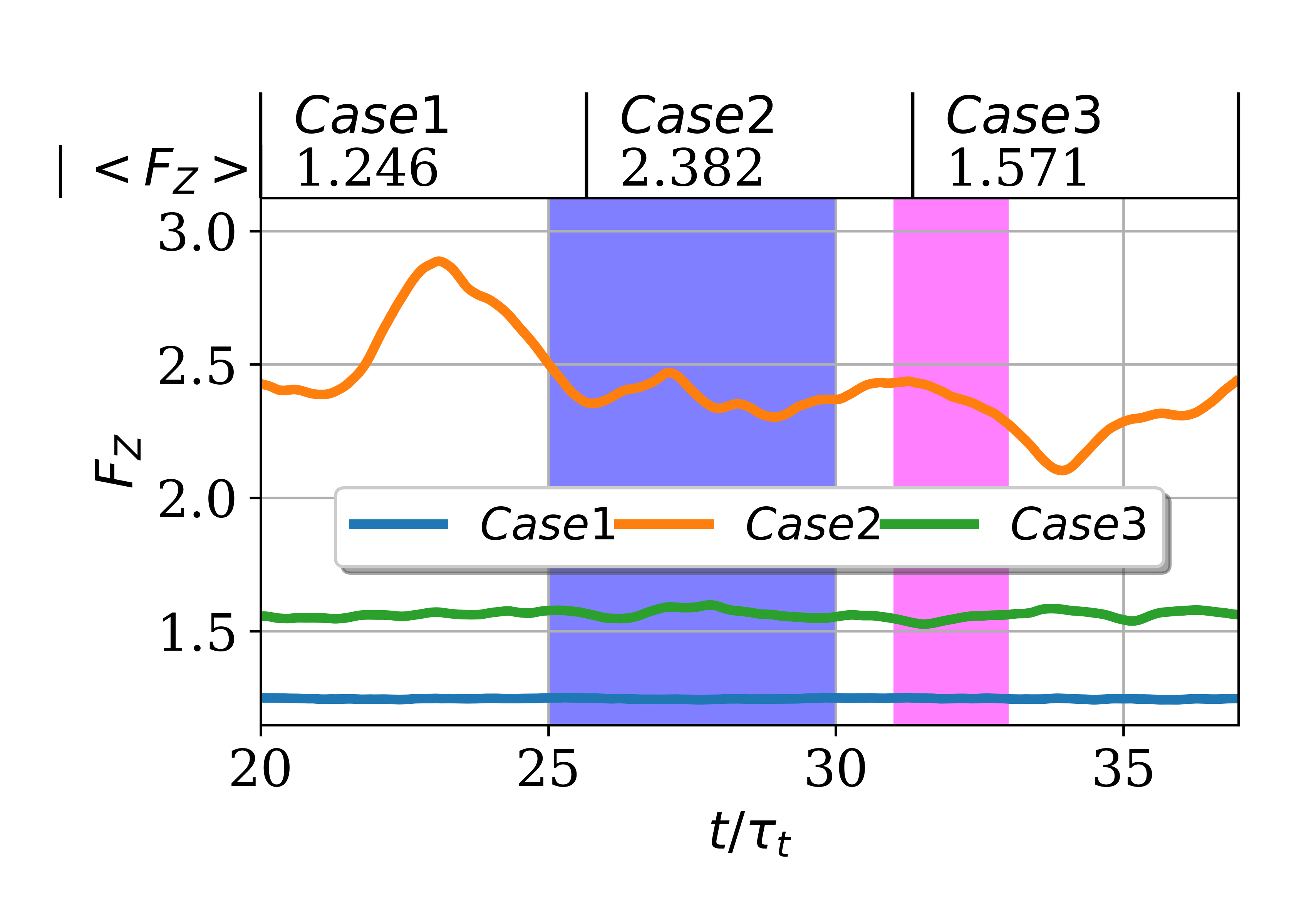}
			\label{fig:FlatnessZ_cases}
		\end{minipage}  
		\vspace{-7mm}
		\caption{The statistics of mixture fraction in each DNS case. \textbf{(Left)} $Z_{rms}$ versus time, \textbf{(right)} the flatness of $Z$ versus time. The time averaged values for each case are given in the figures as well. The color backgrounds are the same as  Fig.\,\ref{fig:ReLambda_vs_time}.   }
		\label{fig:statsZ_vs_time}
	\end{figure}
	It should be noted that at this state turbulent statistics are not exactly constant but fluctuate around a constant value which is typical in DNS with linear forcing~\cite{CarrollP2013_PFa,BassenneM2016_PF}. The scalar statistics of the three DNS cases are shown in Fig.\,\ref{fig:statsZ_vs_time}. We observe that the RMS of mixture fraction is preserved during the training/testing sampling period shown by blue/magenta colors in  Fig.\,\ref{fig:statsZ_vs_time}. Moreover, the fourth moment (flatness) of the scalar is also approximately stationary as shown in Fig.\,\ref{fig:statsZ_vs_time}. The time-averaged values given in the figures indicate that each DNS case shows a different scalar distribution.

	\subsection{Training/testing datasets generation}
	The training/testing datasets are generated by randomly sampling 2D slices of size $256\times 256$  from the DNS databases. After the DNS reached stationary state, 12800 $x-y$ planes are sampled randomly  from 351 3D instances of DNS data in  the time intervall $25 \le t/\tau_{L,0} \le 30$  with discrete sampling times separated by $\Delta t\approx 0.022 \tau_{L,0} \approx 0.5 \tau_{\eta,0}$ for each case and used as training datasets.. One hundred 2D slices from 3D instances in the period encompassing $31 \le t/\tau_{L,0} \le 33$   are randomly selected as test datasets for each case.  
	Physically consistent flipping and rotation~\cite{ChungW2023_CONF} are performed on velocity components channels while the scalar channel is flipped and rotated in a conventional way during training to expand the database and avoid over-fitting. 
	
	The three DNS cases in Table~\ref{tab:sharedSpecInit} are used to perform different numerical numerical experiments. The aim is to assess the generalization capability of the trained networks with respect to scalar distributions.  In \textit{Experiment1} the database with double Pareto distribution of mixture fraction is used as the training database and samples from semi-Gaussian mixture fraction distributions are employed to test the generalization of  the trained network. 
	The reverse approach is used  in \textit{Experiment2} to test the generalization capability of  models  which are trained based on semi-Gaussian distributions. 
	In \textit{Experiment3}  we attempt to improve the results by including both distributions in the training dataset and test whether the trained model can deconvolve samples from both distributions, as well as  out-of-sample fields with  bimodal distribution in \textit{Case3} of Table~\ref{tab:sharedSpecInit}. 
	Note that in \textit{Experiment3}  where the training data is composed of both double Pareto and semi-Gaussian distributions, there are $2\times12800$ images in the training dataset with equal number of samples from each distribution. 
	\begin{table}[!htb]
		\centering
		\caption{Training/testing datasets configurations in different numerical experiments.}
		\label{tab:experiments}       
		\begin{tabular}{p{3cm}p{2cm}p{2cm}p{2cm}}
			\hline\noalign{\smallskip}
			Experiments & Case1   		 &   Case2         &   Case3        \\
			& (doublePareto) &   (semi-Gaussian)    &   (bimodal)   \\
			\hline 
			Experiment1 & Training & Testing & -  \\
			Experiment2 & Testing  & Training & -  \\
			Experiment3 & Training/testing  & Training/testing & Testing \\	
			\noalign{\smallskip}\hline\noalign{\smallskip}
		\end{tabular}
	\end{table}

	\subsection{Training strategy}
	All numerical experiments in Table~\ref{tab:experiments} are performed with a scaling factor of 4 between the low resolution (LR) and high resolution (HR) data.  
	The standardized $256\times 256$ HR training slices containing four input fields i.e. velocity vector components and the mixture fraction, are  randomly cropped on-the-fly to $128\times 128$ input data. The LR images are obtained by filtering and down-sampling the HR data by a 2D  top-hat filter with the width of 4. The batch size is set to 64. The training procedure in each numerical experiment is composed of two steps. First, a pre-training step is performed where the RRDB generator  is trained  with the loss function in Eq.\,\eqref{eq:Lgan}  with $\beta_{\mathcal{L}_{1}} = 0.2$,  $\beta_{\mathcal{L}_{grad}} =0.8$, and  $\beta_{\mathcal{L}_{AD}} = 0$. The learning rate is set to $10^{-4}$ and the ADAM optimizer~\cite{KingmaD2014_3ICLRI2-CTP} with  $\beta_{1,ADAM} = 0.9$, and  $\beta_{2,ADAM} = 0.99$ is used. 
	After the pre-training with 400K iterations, the GAN training including the adversarial loss function  with $\beta_{\mathcal{L}_{1}} = 0.284$,  $\beta_{\mathcal{L}_{grad}} =0.7$ , and $\beta_{\mathcal{L}_{AD}} = 0.016 $  is performed for 100K more iterations. 
	A similar optimizer as for the pre-training step is used. 
	The  learning rates of the discriminator and generator are set to $5\cdot10^{-6}$.   
	The training is performed on a single Nvidia 4090 GPU. Each training takes approximately 96 hours.

	\clearpage
	\section{Results}

	\subsection{Experiment1: Training on double Pareto and testing on semi-Gaussian}
	
	In \textit{Experiment1} the RRDB-GAN  is trained on the dataset from \textit{Case1} with a double Pareto mixture fraction ($Z$) distribution. Both velocity components and mixture fraction fields are super-resolved by a factor of 4.

\begin{figure}[!htb]
		\centering 
		\includegraphics[width=0.9\textwidth,trim=0 3.6cm 0 2.9cm, clip]{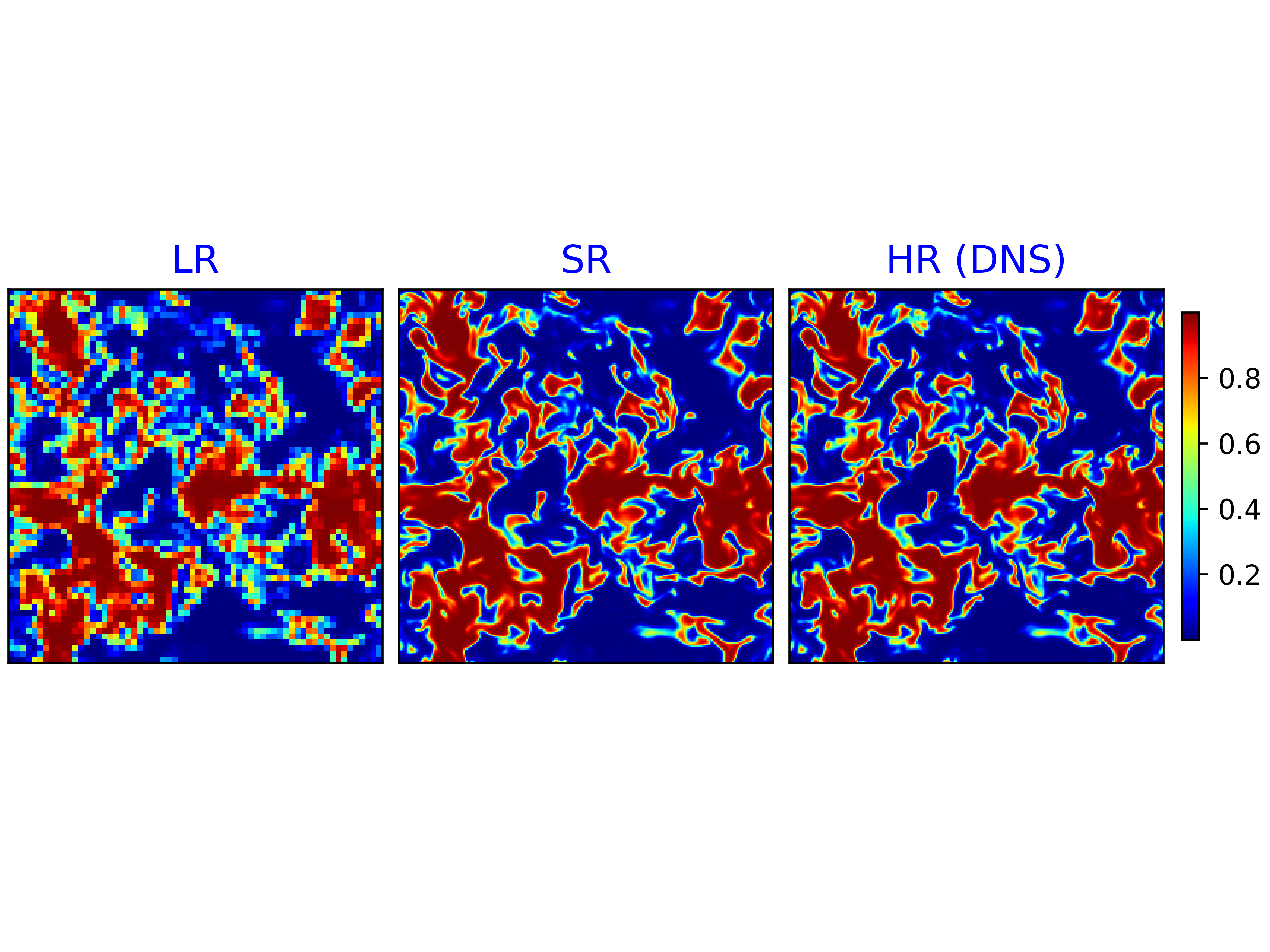}
		\caption{\textit{Experiment1}: Low resolution (LR), super-resolved (SR) and ground-truth (HR) $Z$ field for test data sampled from \textit{Case1} with double Pareto $Z$ distribution. }
		\label{fig:exp1_6141_inference_fdoublePareto_outOfSample_scalar_4x_test4x_31000_img98_full_256}
	\end{figure}

\begin{figure}[!htb]
		\centering
		\begin{minipage}[c]{.45\textwidth}
			\centering
			\includegraphics[width=\textwidth,trim=0 0cm 0 0.8cm, clip]{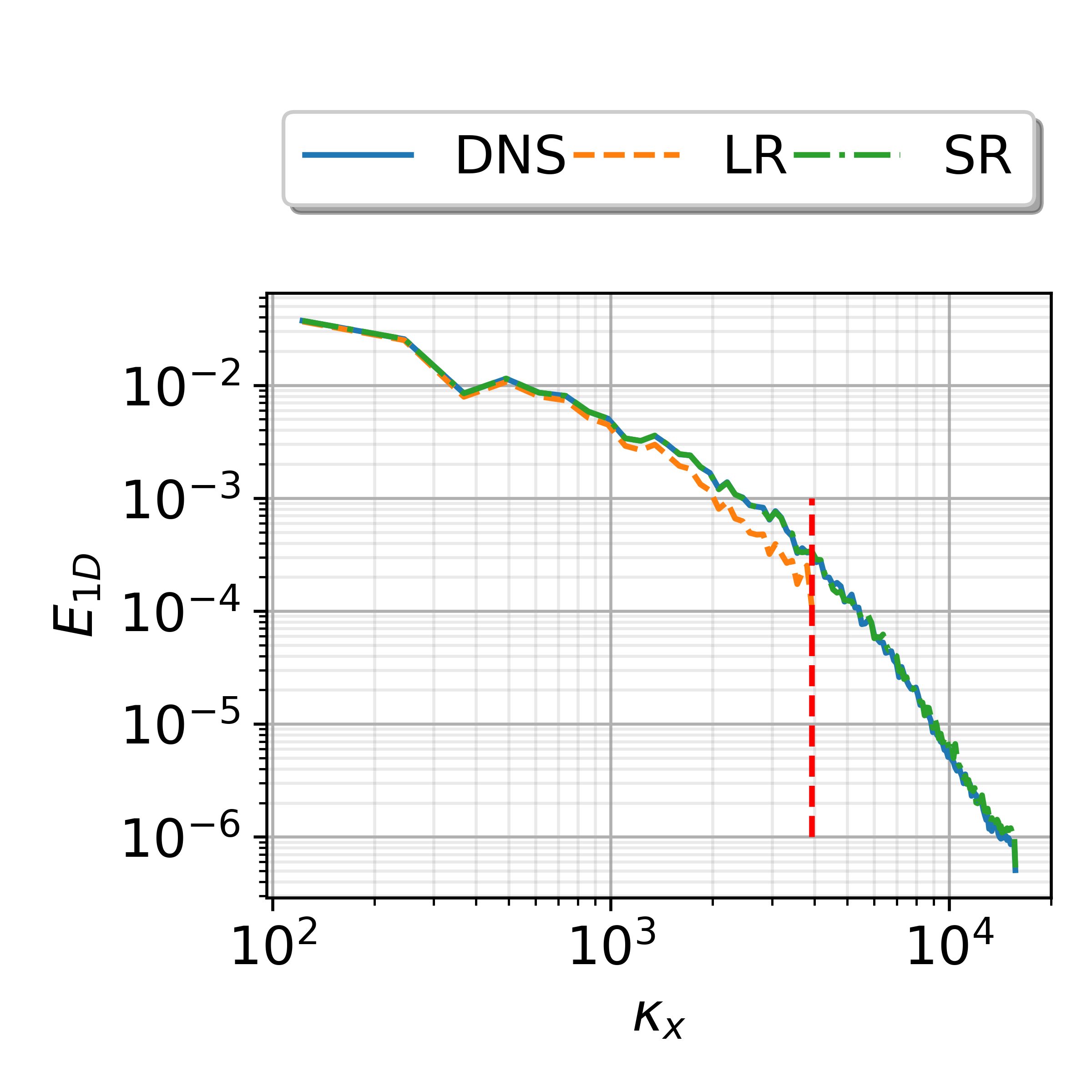}
			\vspace{-7mm}
			\subcaption{}\label{fig:exp1_inference_fdoublePareto_outOfSample_lines_scalarSpec}
		\end{minipage}
		\begin{minipage}[c]{.45\textwidth}
			\centering
			\includegraphics[width=\textwidth,trim=0 0cm 0 1cm, clip]{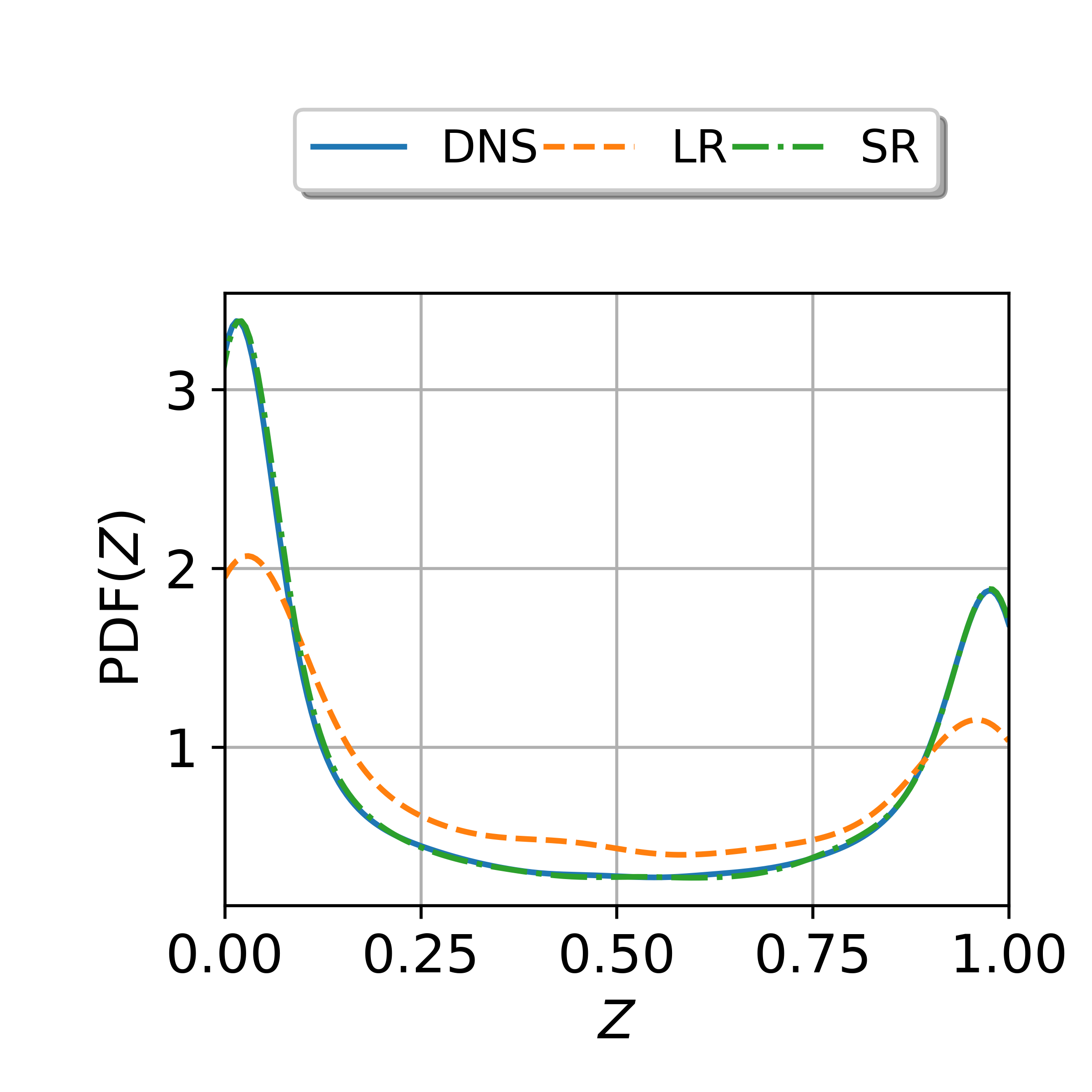}
			\vspace{-7mm}
			\subcaption{}\label{fig:exp1_inference_fdoublePareto_outOfSample_lines_scalarPDF}
		\end{minipage} 
		\vspace{-3mm}
		\caption{\textit{Experiment1}:  1D scalar spectrum and (\textbf{b}) PDF for ground-truth (DNS), filtered DNS (LR), and super-resolved (SR) for  test  data sampled from \textit{Case1} with double Pareto $Z$ distribution. }
		\label{fig:exp1_inference_fdoublePareto_outOfSample_lines}
	\end{figure}

	In Fig.\,\ref{fig:exp1_6141_inference_fdoublePareto_outOfSample_scalar_4x_test4x_31000_img98_full_256} the reconstructed mixture fraction field (SR) from the low-resolution data (LR) is compared with ground-truth (HR) DNS data. 
	All inference data are $x-y$ slices on a random $z$ axis of the DNS domain. They are  sampled from the test dataset and are not seen during the training. Further, they are well separated from the training samples by one turbulent time scale, cf. Fig.\,\ref{fig:ReLambda_vs_time}. 
	The good qualitative agreement between the DNS and SR fields is quantitatively assessed with the help of   
	Fig.\,\ref{fig:exp1_inference_fdoublePareto_outOfSample_lines} where the scalar 1D  spectra and PDFs are shown. It can be seen that both quantities are well reconstructed. Specifically, in Fig.\,\ref{fig:exp1_inference_fdoublePareto_outOfSample_lines_scalarSpec} the under-resolved (left side of the cut-off denoted by vertical red line) and unresolved (right side of the cut-off) are well reconstructed. 

\begin{figure}[!htb]
		\centering
		\begin{minipage}[c]{.45\textwidth}
			\centering
			\includegraphics[width=\textwidth,trim=0 0cm 0 0.8cm, clip]{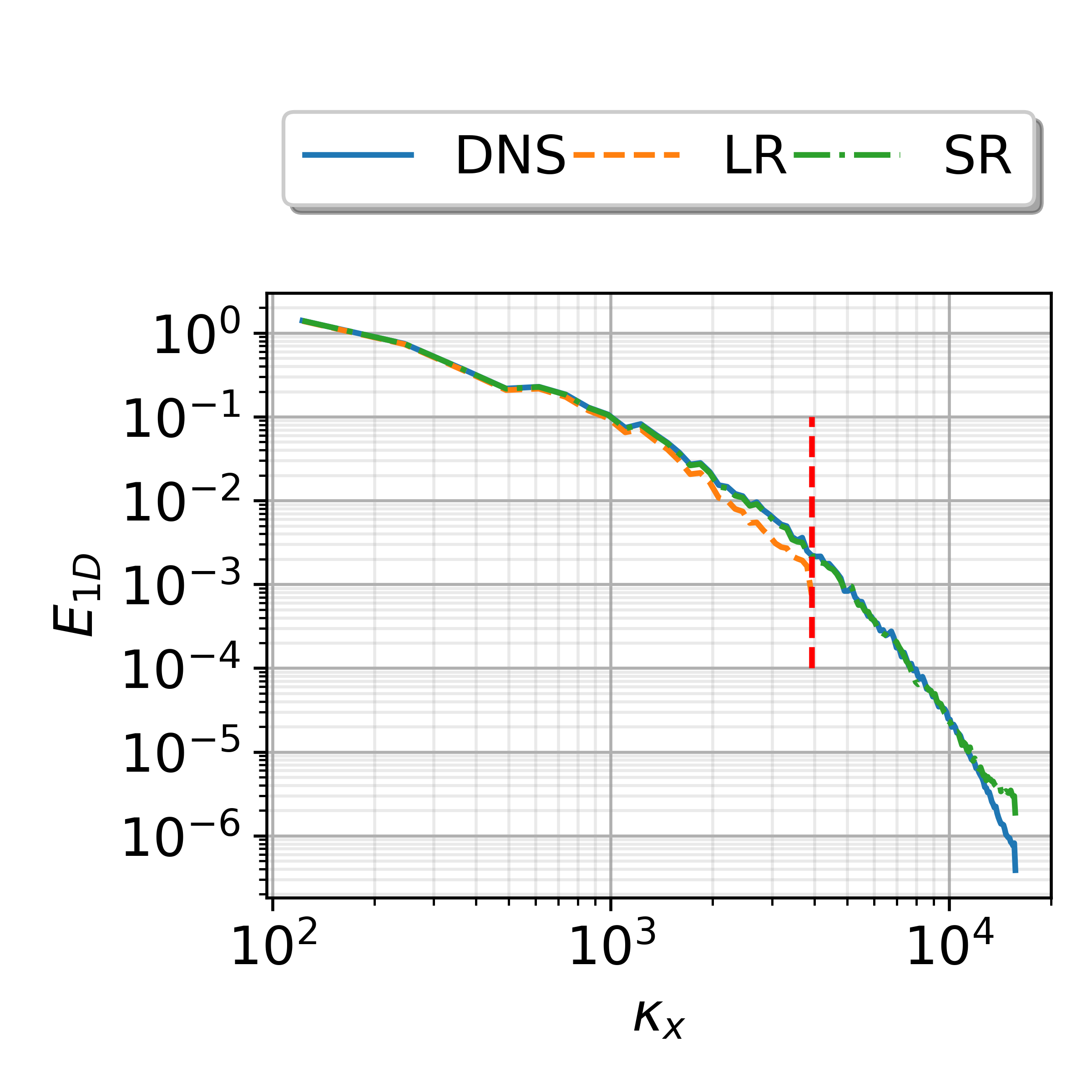}
			\vspace{-7mm}
			\subcaption{}\label{fig:exp1_inference_fdoublePareto_outOfSample_lines_velSpec}
		\end{minipage}
		\begin{minipage}[c]{.45\textwidth}
			\centering
			\includegraphics[width=\textwidth,trim=0 0cm 0 1cm, clip]{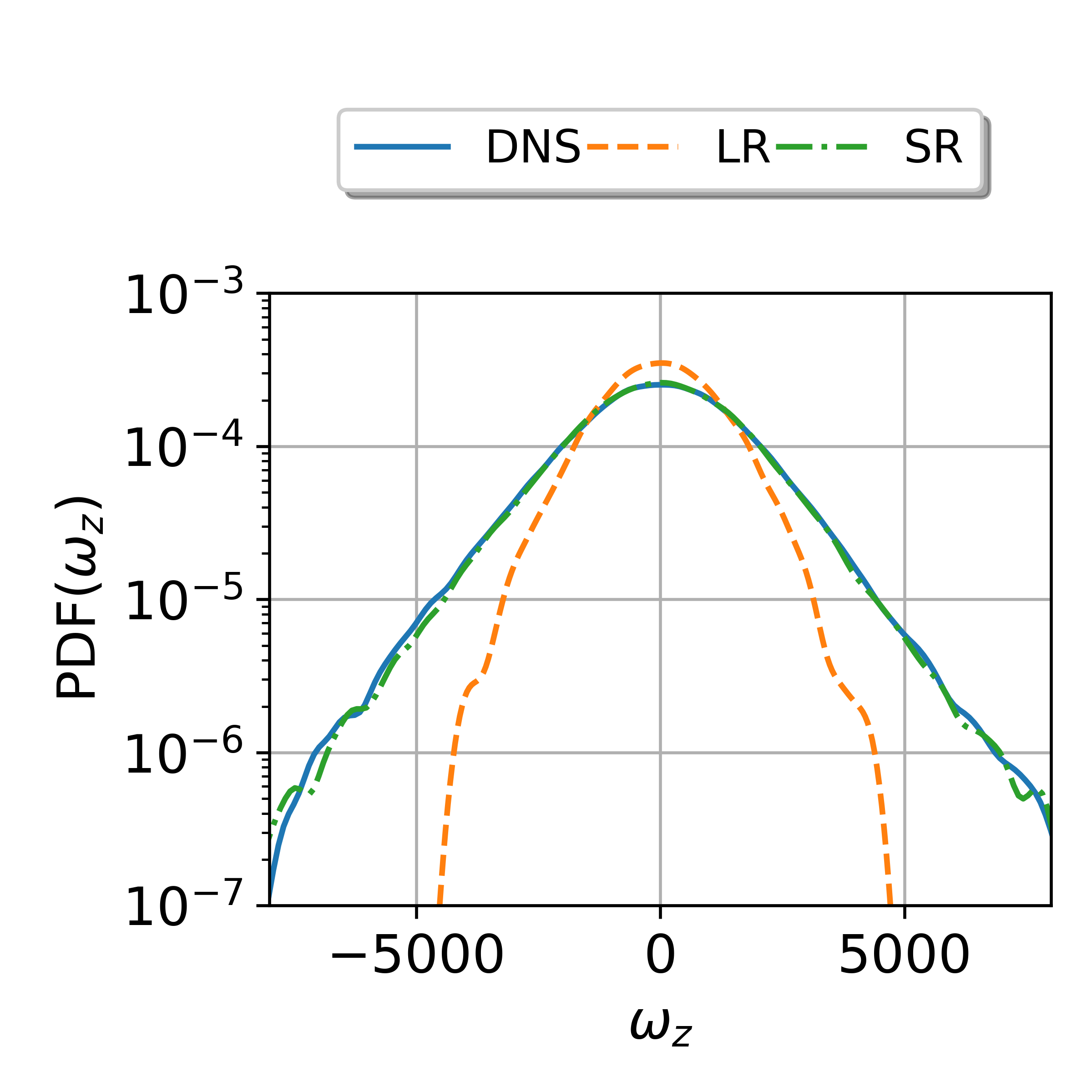}
			\vspace{-7mm}
			\subcaption{}\label{fig:exp1_inference_fdoublePareto_outOfSample_lines_velPDF}
		\end{minipage}
		\vspace{-3mm}
		\caption{\textit{Experiment1}:  1D velocity energy spectrum and (\textbf{b}) PDF of $z$ component of vorticity  for ground-truth (DNS), filtered DNS (LR), and super-resolved (SR) for  test  data sampled from \textit{Case1}. }
		\label{fig:exp1_inference_fdoublePareto_outOfSample_lines_vel}
	\end{figure}

	Three velocity components and mixture fraction are super-resolved simultaneously by the model. For the sake of brevity only a quantitative comparison of the velocity energy spectra and PDF of the $z$-component  of the vorticity vector are presented in Fig.\,\ref{fig:exp1_inference_fdoublePareto_outOfSample_lines_vel} and demonstrate a good agreement with the target DNS data. 
	
\begin{figure}[!htb] 
		\centering
		\begin{minipage}[c]{0.9\textwidth}
			\centering
			\includegraphics[width=\textwidth,trim=0 3.6cm 0 2.9cm, clip]{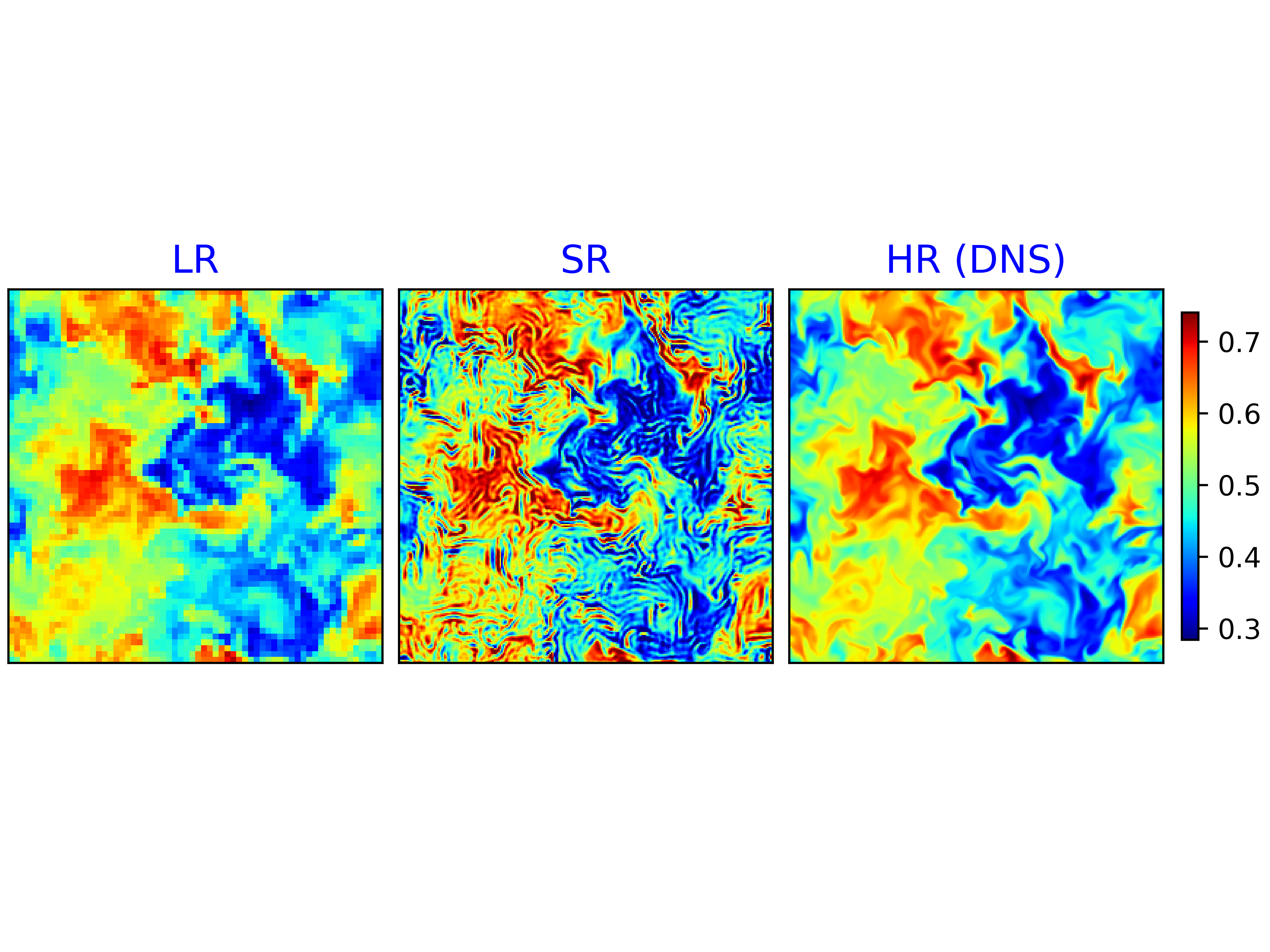}
		\end{minipage}
		\caption{\textit{Experiment1}: Contours of $Z$ field for  test data sampled from \textit{Case2} with  semi-Gaussian $Z$ distribution. }
		\label{fig:exp1_inference_frms010Gaussian_Generalization_outOfSample_lines_scalarimg}
	\end{figure}

	The generalization capability of the trained model is tested in Fig.\,\ref{fig:exp1_inference_frms010Gaussian_Generalization_outOfSample_lines_scalarimg} where we test the model which is trained on double Pareto $Z$ distributions to super-resolve the fields which are taken from semi-Gaussian $Z$ distributions of \textit{Case2}. It can be seen in Fig.\,\ref{fig:exp1_inference_frms010Gaussian_Generalization_outOfSample_lines_scalarimg} that the models fail by generating nonphysical waves and structures. 
	Further inspection (not shown for the sake of brevity) shows that changing the  normalization factors of the test data i.e. normalizing by  mean and standard  deviation of the test  dataset instead of the training one does not improve the results. 

	\clearpage
	\subsection{Experiment2: Training on semi-Gaussian  and testing on double Pareto}
\begin{figure}[!htb] 
		\centering
		\includegraphics[width=0.9\textwidth,trim=0 3.7cm 0 2.9cm, clip]{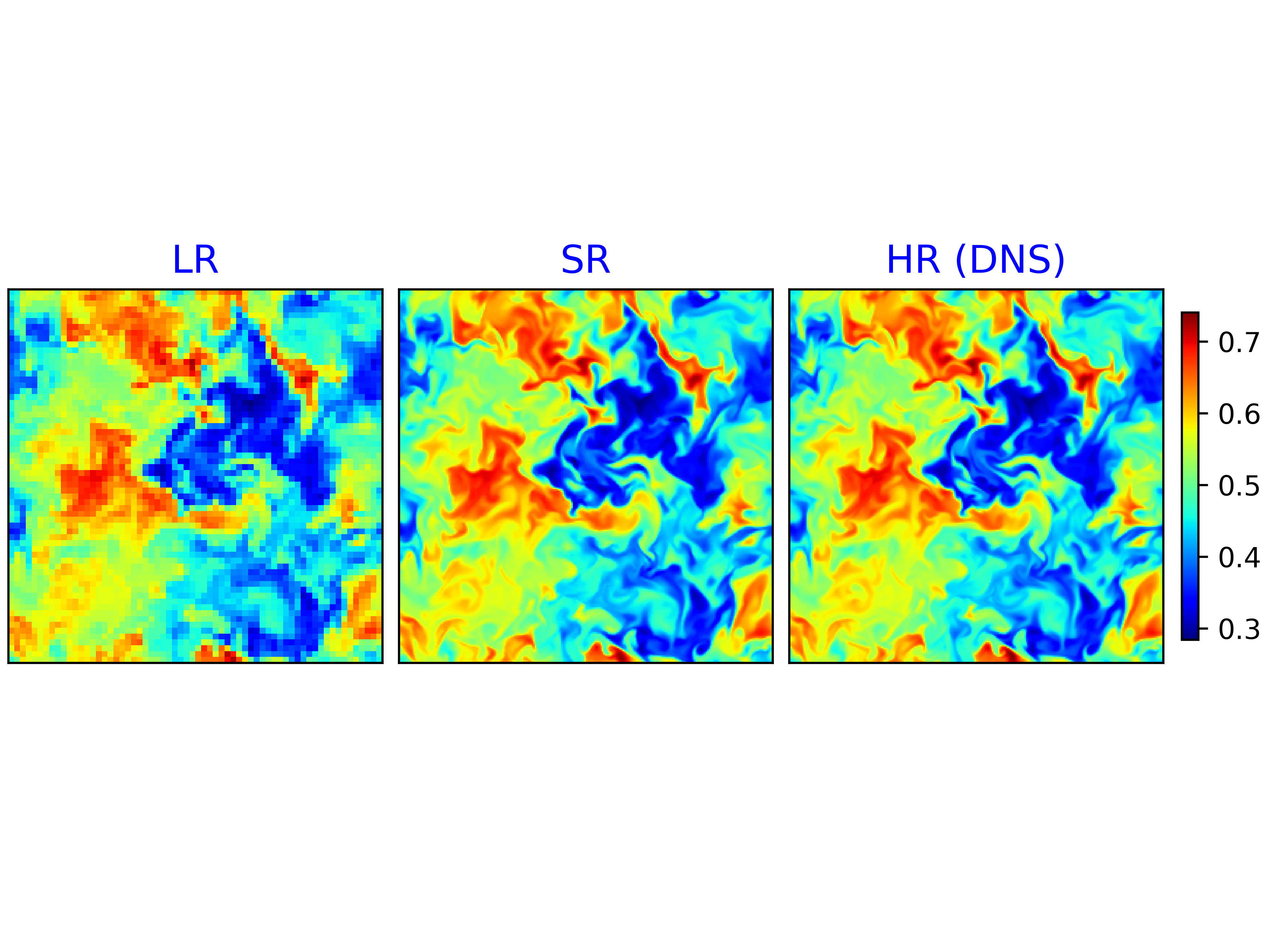}
		\caption{\textit{Experiment2}: Low resolution (LR), super-resolved (SR) and ground-truth (HR) $Z$ field for  test  data sampled from \textit{Case2} with semi-Gaussian $Z$ distribution. }
		\label{fig:exp2_inference_frms010Gaussian_outOfSample_imgs_z}
	\end{figure}

\begin{figure}[!htb]
		\centering 
		\begin{minipage}[c]{0.45\textwidth}
			\centering
			\includegraphics[width=\textwidth,trim=0.3cm 0 0.3cm 1cm, clip]{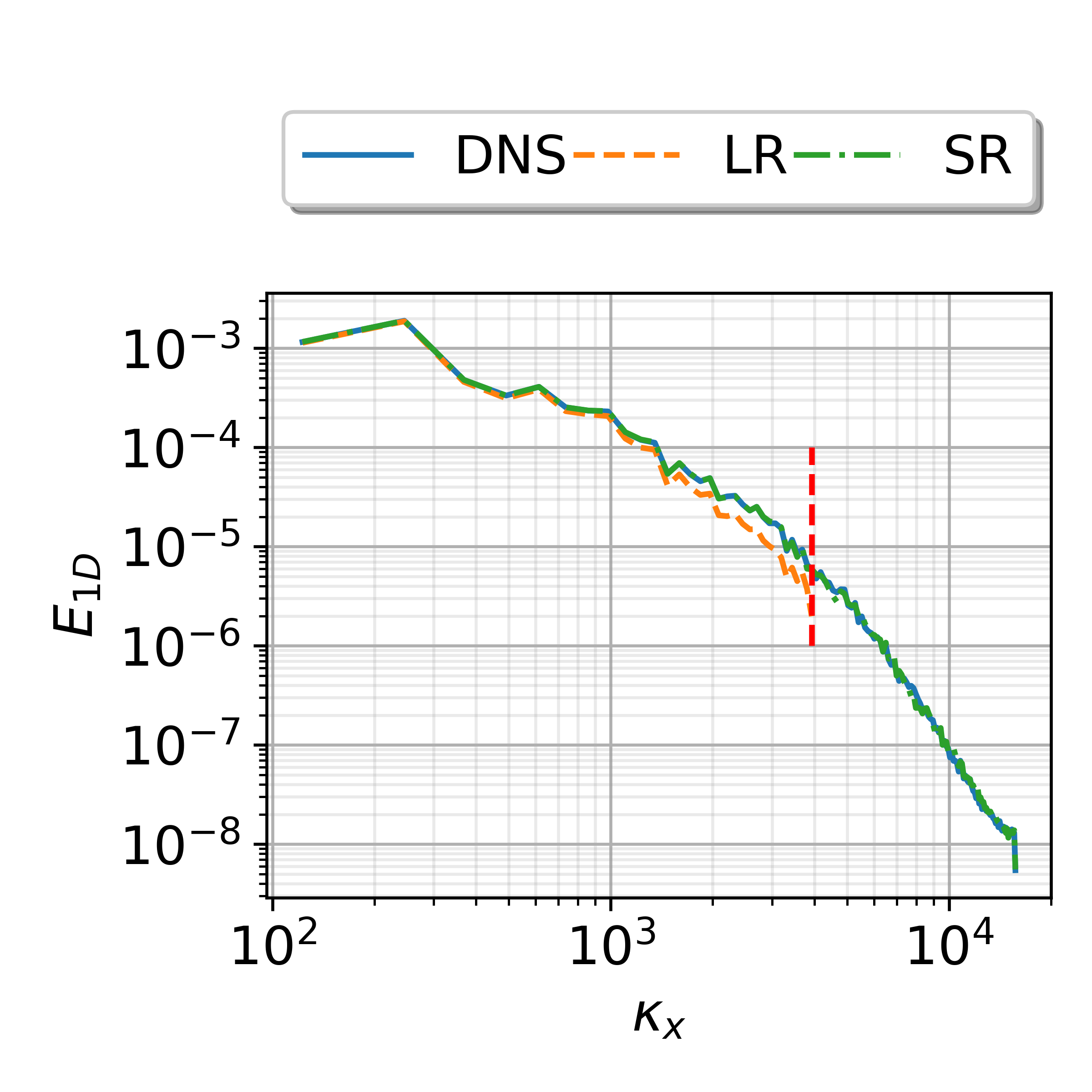}
			\vspace{-7mm}
			\subcaption{}
			\label{fig:exp2_inference_frms010Gaussian_outOfSample_scalarSpec}
		\end{minipage}
		\begin{minipage}[c]{0.45\textwidth}
			\centering
			\includegraphics[width=\textwidth,trim=0.3cm 0 0.3cm 1cm, clip]{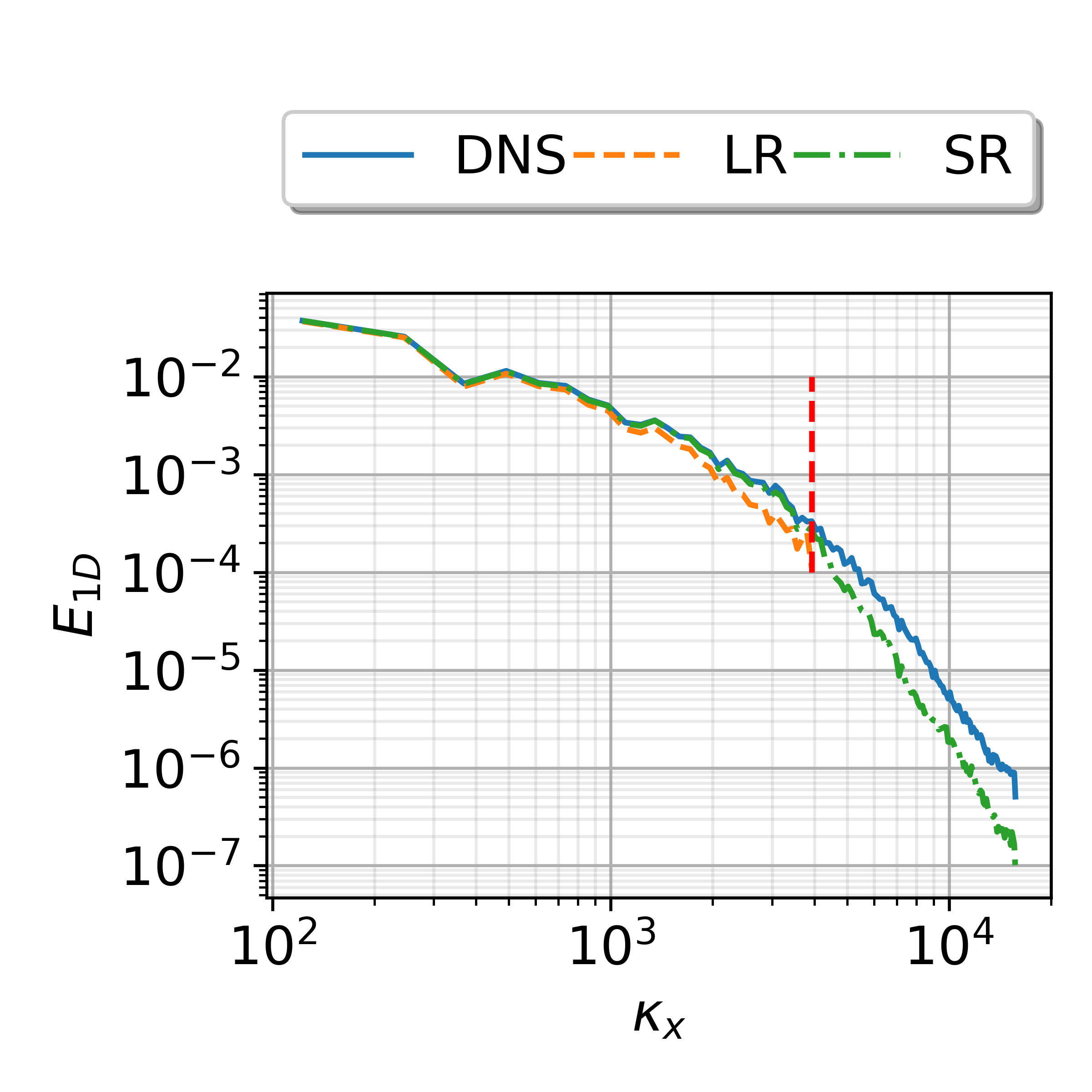}
			\vspace{-7mm}
			\subcaption{}
			\label{fig:exp2_inference_fdoublePareto_Generalization_outOfSample_lines_scalarSpec}
		\end{minipage} 	
		\vspace{-3mm}
		\caption{\textit{Experiment2}: The scalar spectra for  test data sampled from (\textbf{a})  \textit{Case2} with  semi-Gaussian $Z$ distribution and (\textbf{b})  \textit{Case1} with  double Pareto $Z$ distribution. }
		\label{fig:exp2_inference_frms010Gaussian_outOfSample_and_inference_fdoublePareto_Generalization_outOfSample_lines}
	\end{figure}

	In this section the results of  \textit{Experiment2} are presented in which  the RRDB-GAN  is trained on the dataset from \textit{Case2} with semi-Gaussian $Z$ distribution. The same test data as in Fig.\,\ref{fig:exp1_inference_frms010Gaussian_Generalization_outOfSample_lines_scalarimg} is used and shown in  Fig.\,\ref{fig:exp2_inference_frms010Gaussian_outOfSample_imgs_z}. As can be seen in Fig.\,\ref{fig:exp2_inference_frms010Gaussian_outOfSample_imgs_z}   when the network is trained on the same distribution as the testing distributions it can  reconstruct the field  rather well. The scalar spectra are compared in Fig.\,\ref{fig:exp2_inference_frms010Gaussian_outOfSample_scalarSpec} showing the success of the network in super-resolution of the field. 
	In Fig.\,\ref{fig:exp2_inference_fdoublePareto_Generalization_outOfSample_lines_scalarSpec} the generalization capability of the trained model in \textit{Experiment2} is assessed by using a test sample from a $Z$ distribution which is not seen during the training i.e., double Pareto $Z$ distribution. While the results are better compared to the previous generalization test in \textit{Experiment1} the model still largely under predicts  $Z$ fluctuations after the cut-off. Some discrepancies can be also observed near the cut-off. We can conclude from these results that the $Z$ distribution has an effect on the generalization capability  of the super-resolution model.  
	In the next section we try to test if a generalized model can be trained with the capability of super-resolving different $Z$ distributions. 

	\clearpage
	\subsection{Experiment3: Training on semi-Gaussian  and  double Pareto distributions and testing on bimodal}

\begin{figure}[!htb] 
		\begin{minipage}[c]{\textwidth}
			\centering
			\includegraphics[width=0.7\textwidth,trim=0 3.5cm 0 2.5cm, clip]{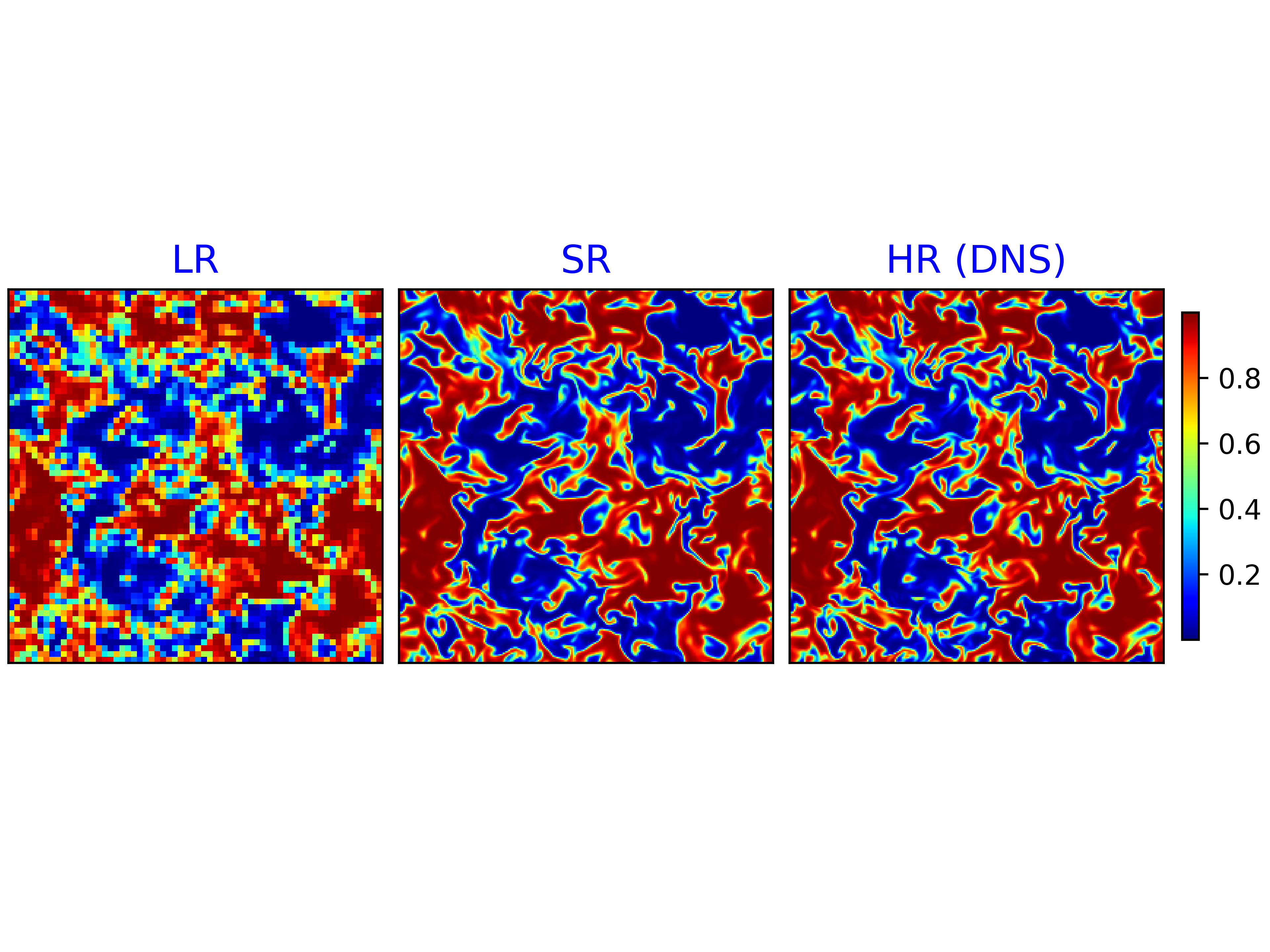}
		\end{minipage}
		\begin{minipage}[c]{\textwidth}
			\centering
			\includegraphics[width=0.7\textwidth,trim=0 3.5cm 0.0cm 3.7cm, clip]{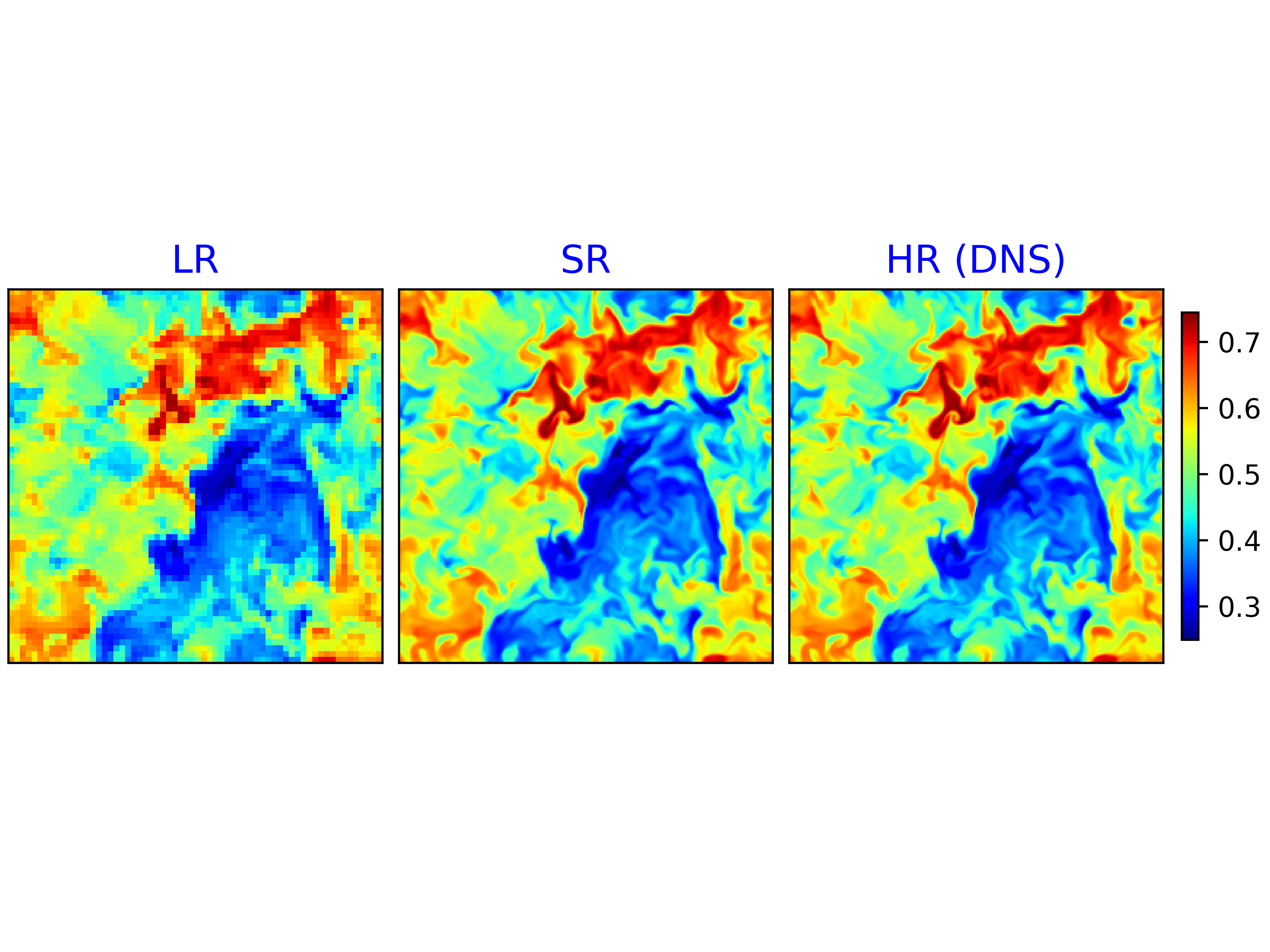}
		\end{minipage}
		\begin{minipage}[c]{\textwidth}
			\centering
			\includegraphics[width=0.7\textwidth,trim=0 3.6cm 0cm 3.7cm, clip]{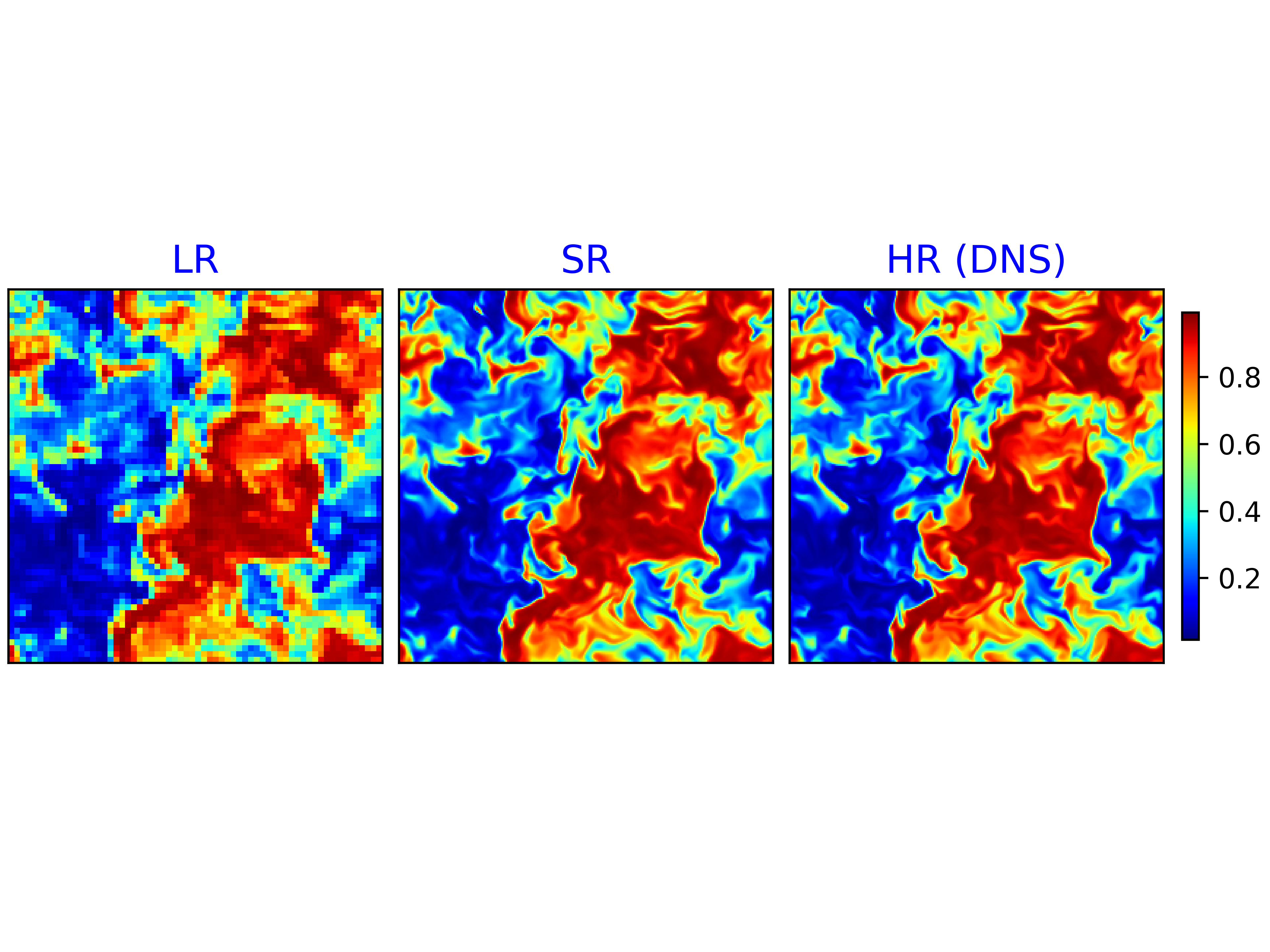}
		\end{minipage}   
		\caption{\textit{Experiment3}: Super-resolution of $Z$ field using a model that is trained on  combined \textit{Case1} and \textit{Case2} datasets with double Pareto and semi-Gaussian $Z$ distributions. Test data are sampled from (\textbf{first row}) \textit{Case1}, (\textbf{second row}) \textit{Case2}, and (\textbf{last row}) \textit{Case3}, each with a unique scalar distribution.  }
		\label{fig:exp3_inference_imgs_scalar}
	\end{figure}

	In \textit{Experiment3} we use a training dataset which is composed of the two extreme $Z$ distributions i.e., double Pareto and semi-Gaussian distributions (cf. Table~\ref{tab:experiments}). In Fig.\,\ref{fig:exp3_inference_imgs_scalar}
	\begin{figure}[!htb] 
		\begin{minipage}[c]{0.33\textwidth}
			\centering
			\includegraphics[width=\textwidth,trim=0.5cm 0 0.3cm 1cm, clip]{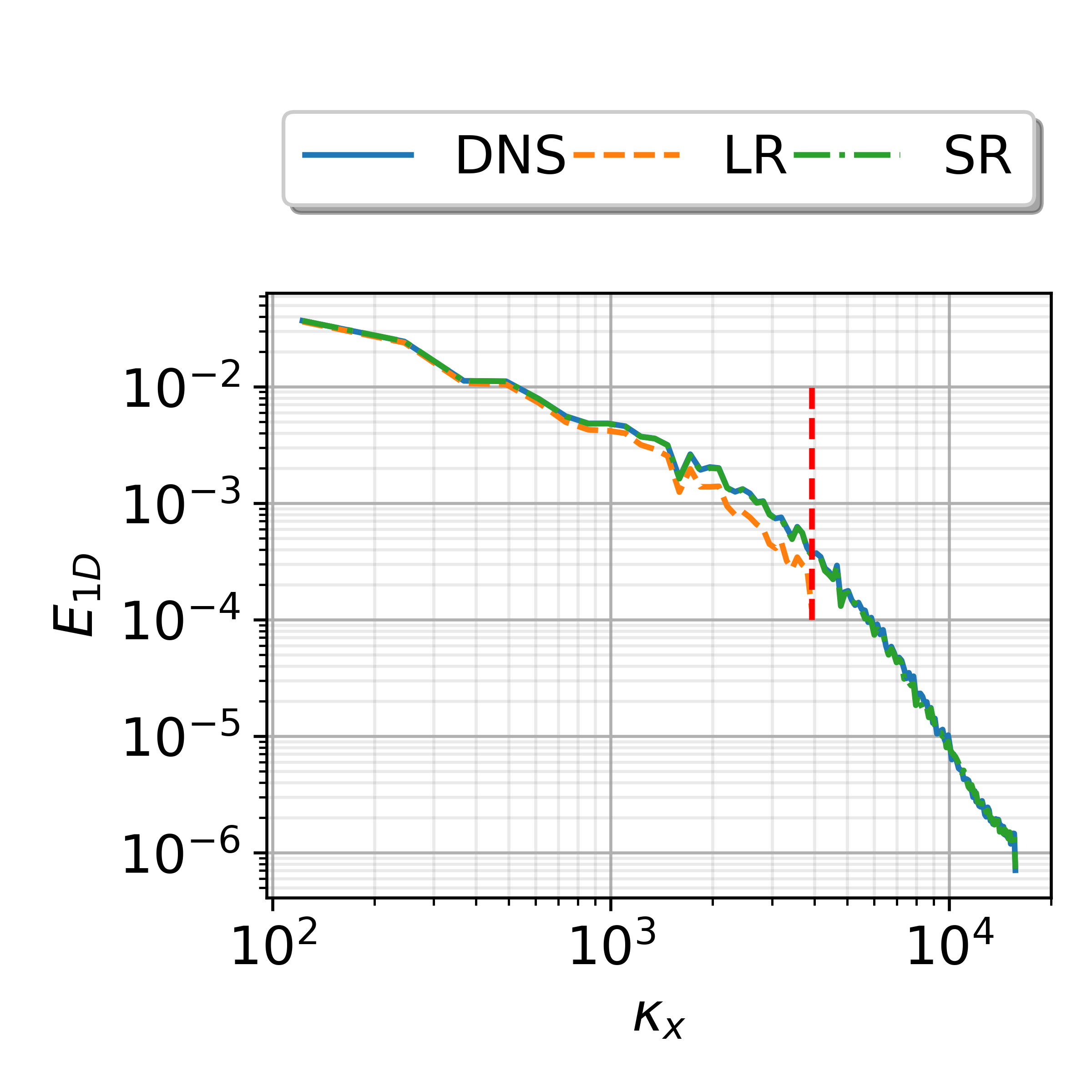}
			\vspace{-7mm}
			\subcaption{}
		\end{minipage}
		\begin{minipage}[c]{0.33\textwidth}
			\centering
			\includegraphics[width=\textwidth,trim=0.5cm 0 0.3cm 1cm, clip]{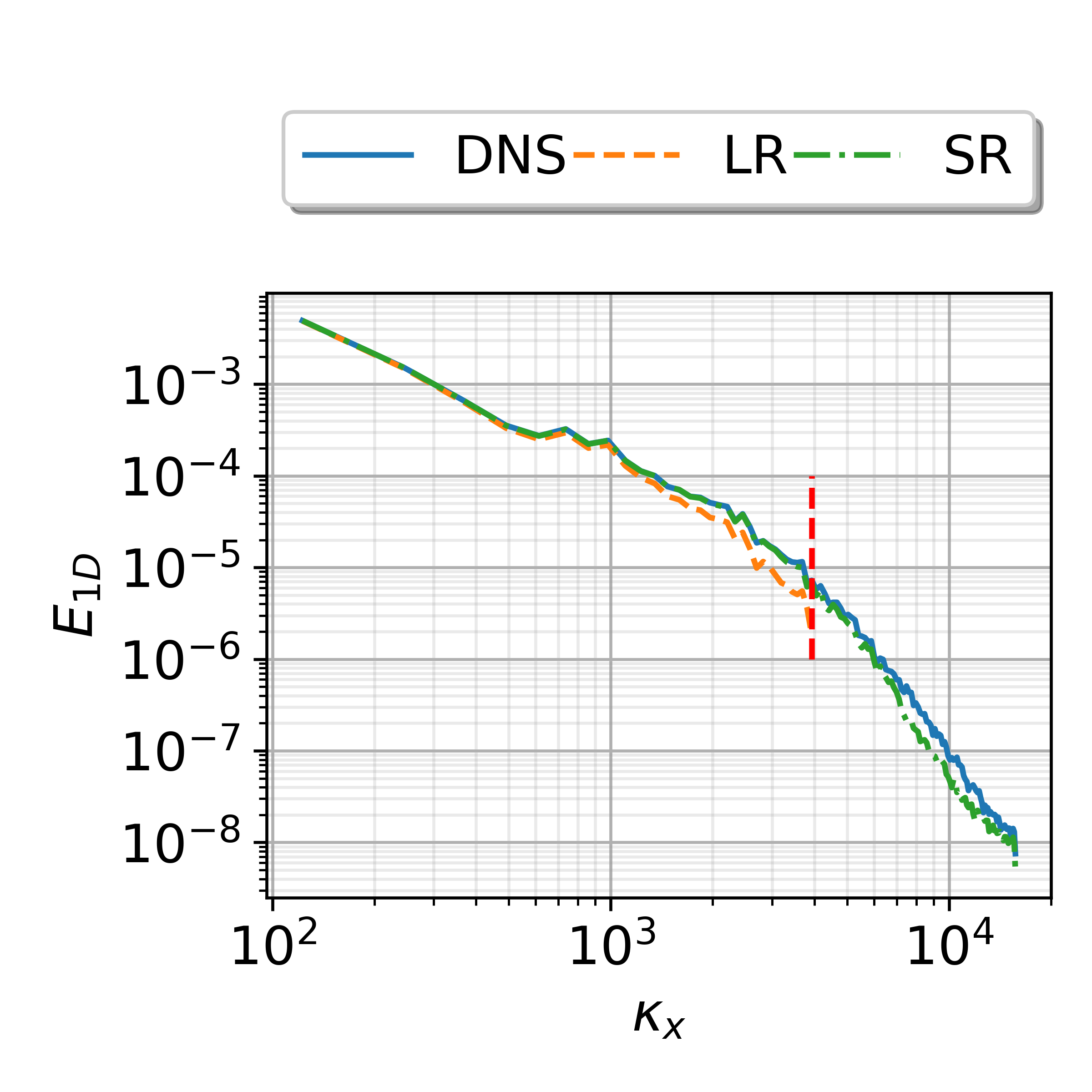}
			\vspace{-7mm}
			\subcaption{}
			\label{fig:exp3_62021_inference_frms010Gaussian_outOfSample_scalar_E_1D_4x_test4x_738000_img95}
		\end{minipage}
		\begin{minipage}[c]{0.33\textwidth}
			\centering
			\includegraphics[width=\textwidth,trim=0.5cm 0 0.3cm 1cm, clip]{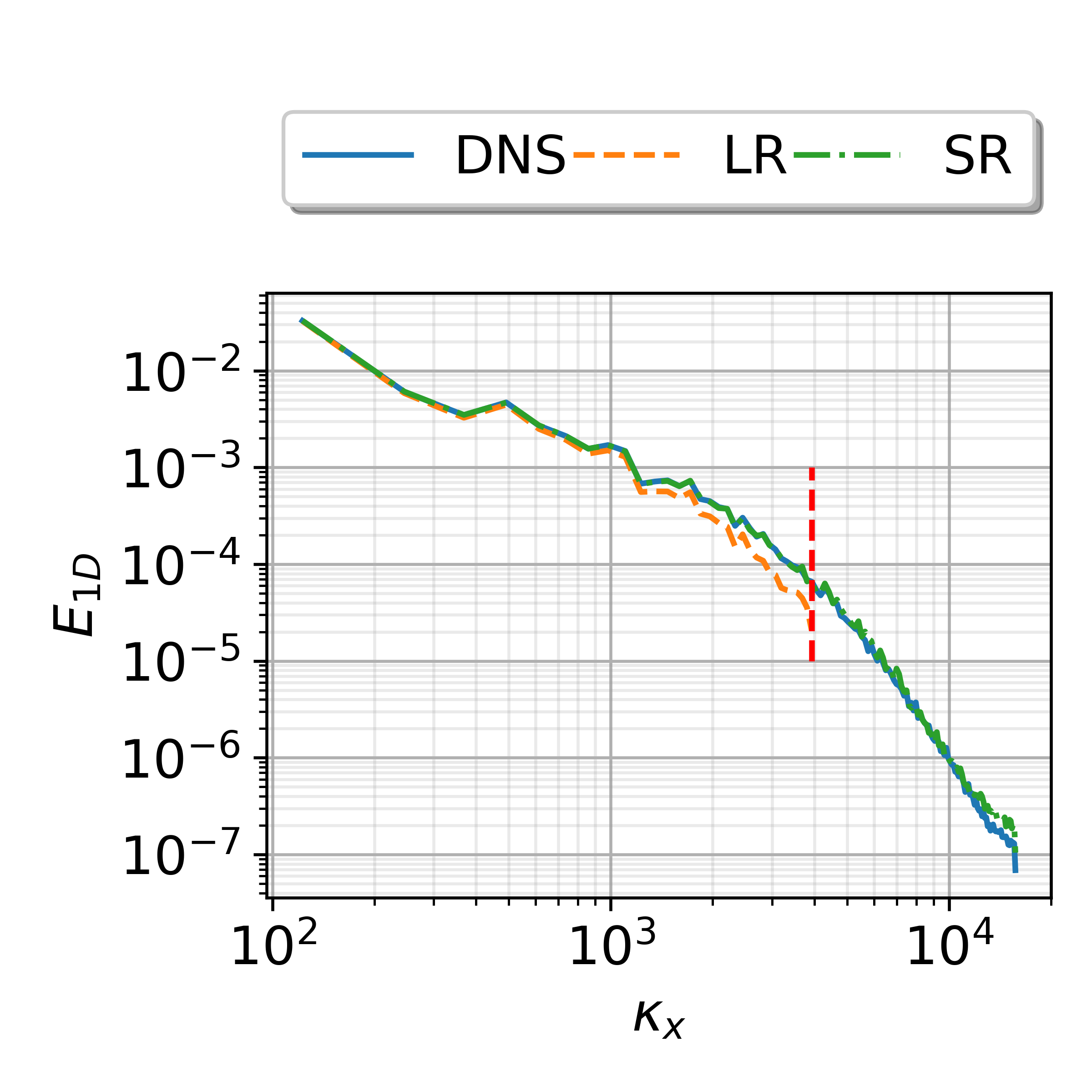}
			\vspace{-7mm}
			\subcaption{}
		\end{minipage}   
		\vspace{-3mm}
		\caption{\textit{Experiment3}: Comparison of scalar spectra. Test data are sampled from (\textbf{a}) \textit{Case1}, (\textbf{b}) \textit{Case2}, and (\textbf{c}) \textit{Case3}, each with a unique scalar distribution.  }
		\label{fig:exp3_inference_scalarSpectra}
	\end{figure}
	 the super-resolved fields are compared with their low and high resolution counterparts. It can be seen that the network is now able to reconstruct unseen samples from both distributions as well as the unseen data from bimodal distributions of \textit{Case3} which were not included in the training dataset.

	\begin{figure}[!htb] 
		\begin{minipage}[c]{0.33\textwidth}
			\centering
			\includegraphics[width=\textwidth,trim=0 0cm 0.5cm 0cm, clip]{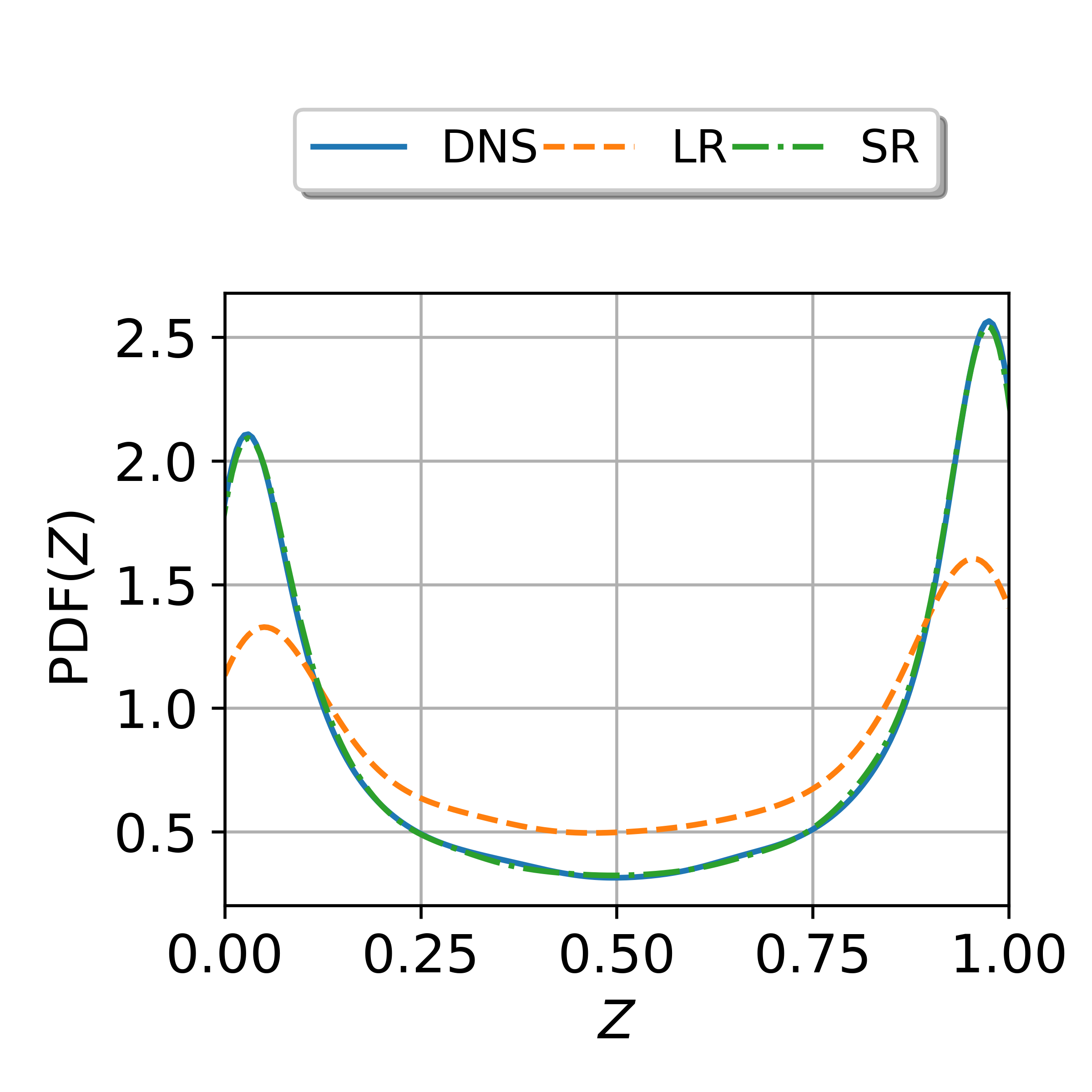}
			\vspace{-7mm}
			\subcaption{}
		\end{minipage}
		\begin{minipage}[c]{0.33\textwidth}
			\centering
			\includegraphics[width=\textwidth,trim=0 0cm 0.5cm 0cm, clip]{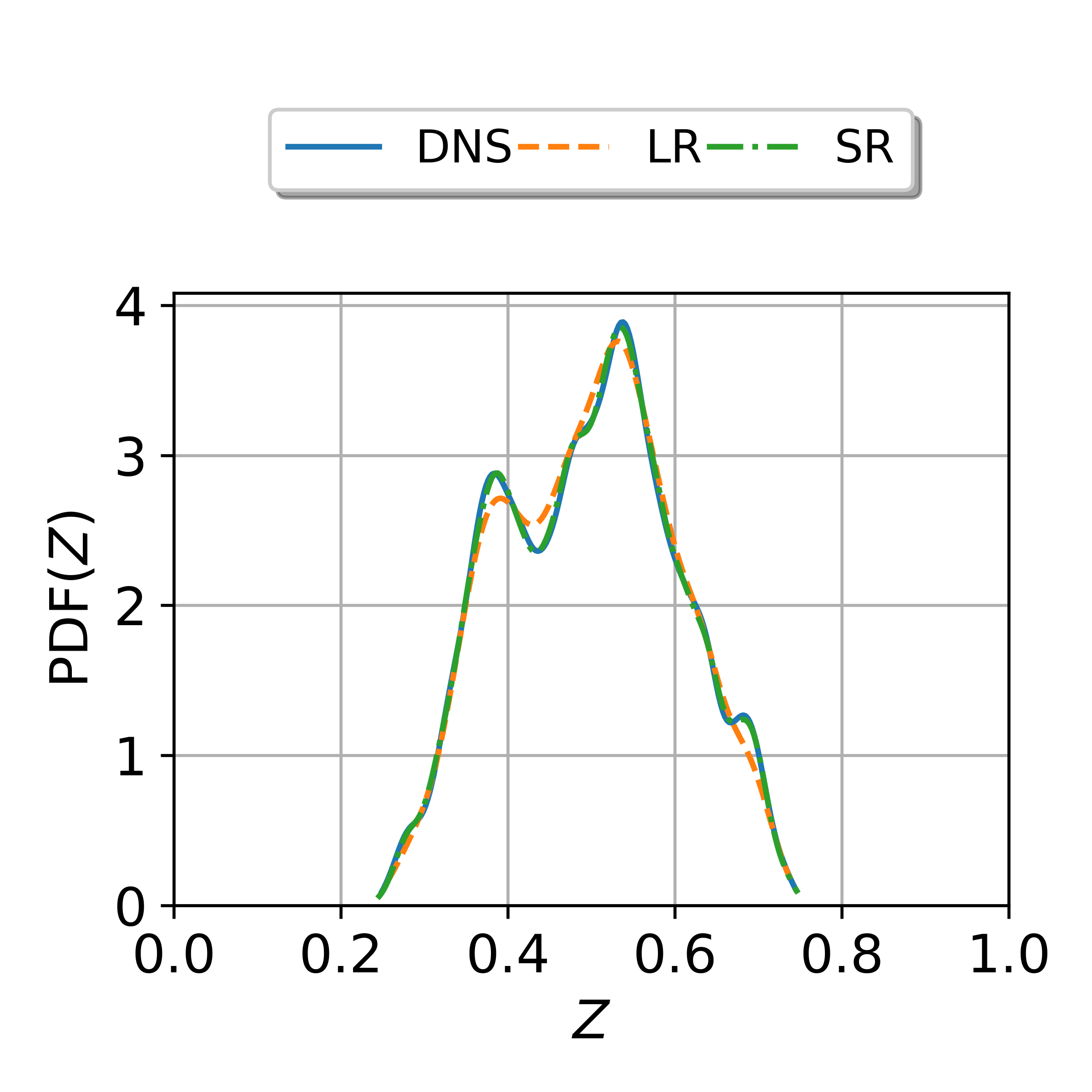}
			\vspace{-7mm}
			\subcaption{}
		\end{minipage}
		\begin{minipage}[c]{0.33\textwidth}
			\centering
			\includegraphics[width=\textwidth,trim=0 0cm 0.5cm 0cm, clip]{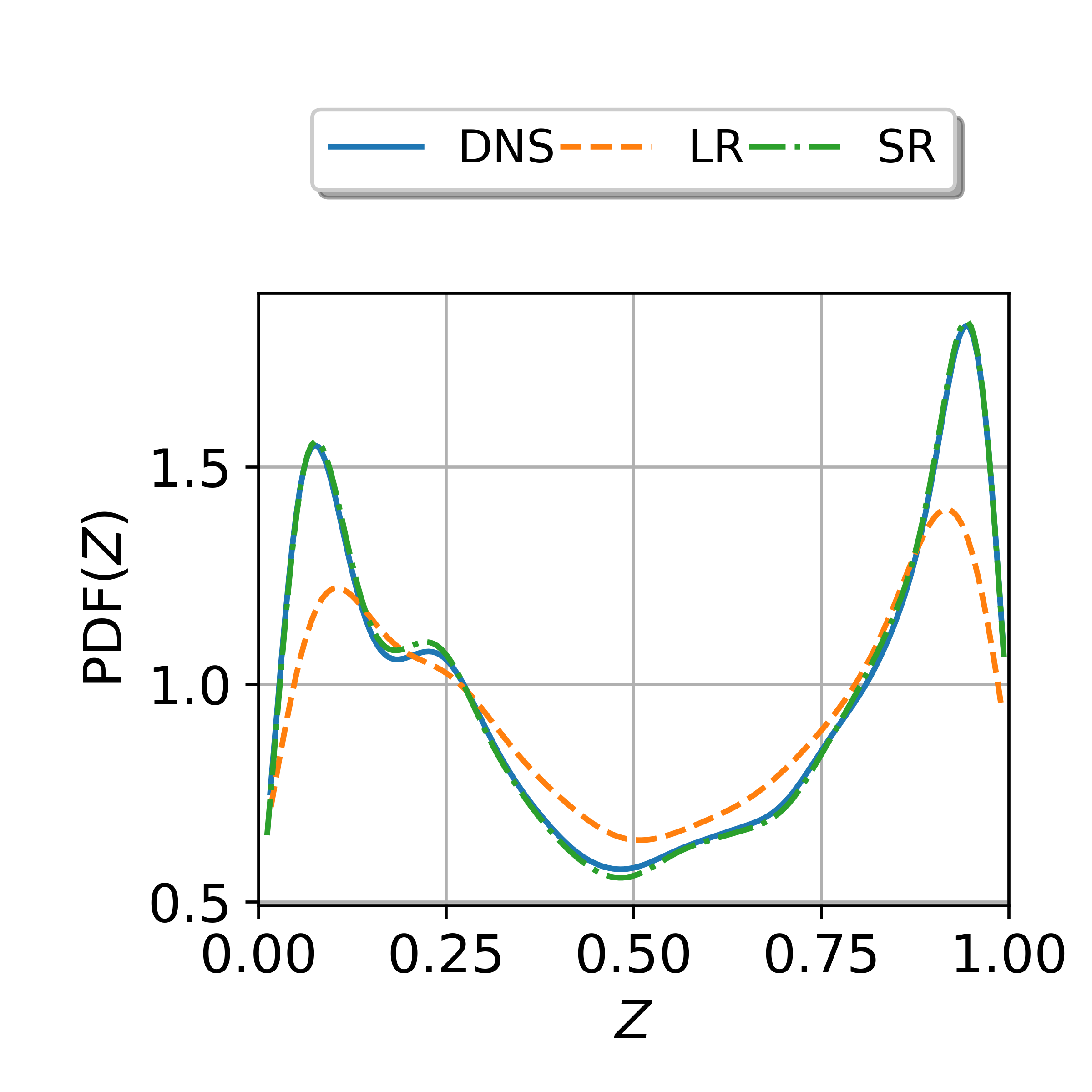}
			\vspace{-7mm}
			\subcaption{}
		\end{minipage}   
		\vspace{-3mm}
		\caption{\textit{Experiment3}: Comparison of scalar PDF. Test data are sampled from (\textbf{a}) \textit{Case1}, (\textbf{b}) \textit{Case2}, and (\textbf{c}) \textit{Case3}, each with a unique scalar distribution.  }
		\label{fig:exp3_inference_scalarPDF}
	\end{figure}
	
	A quantitative comparison is shown in Fig.\,\ref{fig:exp3_inference_scalarSpectra} and Fig.\,\ref{fig:exp3_inference_scalarPDF} where scalar spectra and PDFs of the SR fields are compared with the LR and HR counterparts for different $Z$ distributions. 
	Despite the existence of some under predictions at high wave numbers,    Fig.\,\ref{fig:exp3_62021_inference_frms010Gaussian_outOfSample_scalar_E_1D_4x_test4x_738000_img95} shows that  training with two extreme distributions strongly improves the generalization capability of the model.

	\section{Conclusions}
	In this study super-resolution reconstruction of turbulent velocity fields and a passive scalar (mixture fraction) is performed using a RRDB-GAN. 
	Three different forced turbulence DNS databases each with a different  scalar distribution, namely double Pareto, semi-Gaussian, and bimodal distributions,  have been employed as training/testing datasets. 
	The top-hat filtered and down-sampled DNS data with a filter width of 4 are used as the input of the network to perform  4x super-resolution.  
	The capability of the RRDB-GAN  in generalization to unseen scalar distributions is assessed by performing three  numerical experiments in which training and testing databases with different scalar distributions  are altered. 
	The performance of the model is assessed qualitatively and quantitatively by comparing the generated fields and their statistics with the ground-truth DNS data. 
	It is shown that while the velocity vector field can be reconstructed well, the performance of the super-resolution model in the reconstruction of mixture fraction is affected by the scalar distribution used in the training. 
	The model fails to  super-resolve the out-of-sample scalar distributions which are not included in the training. 
	However, including the two extreme mixture fraction distributions, i.e., double Pareto and semi-Gaussian distributions in the training dataset, significantly improves the performance of the model in super-resolving mixture fraction  fields from unseen test data with double Pareto and semi-Gaussian distributions as well as an unseen bimodal distribution.

	\section{Acknowledgments}
	The authors acknowledge the financial support by the German Research Foundation (DFG) project  No. 513858356.  
	O. T. Stein  acknowledges the financial support by the Helmholtz Association of German Research Centers (HGF), within the research field \textit{Energy}, program \textit{Materials and Technologies for the Energy Transition} (MTET), topic \textit{Resource and Energy Efficiency}. 
	
	\bibliographystyle{unsrt} 
	\bibliography{MULTIPHASE-HLRSJahresbericht-2024}
\end{document}